\documentclass{aa}
\usepackage[varg]{txfonts}
\usepackage{natbib}
\usepackage{graphicx}
\usepackage{amsmath}
\usepackage{amssymb}
\usepackage{xcolor}
\usepackage[normalem]{ulem}
\usepackage{hyperref}
\bibpunct{(}{)}{;}{a}{}{,} 
\begin{document}
\title{Third order correlations and skewness in convection.}
\subtitle{{I}. A new approach suitable for three-equation
        non-local models.}

\author{F.\ Kupka\inst{1,2,3}
}
\institute{Faculty Comp. Science and Appl. Math., Univ. of Applied Sciences, Technikum Wien, H\"ochst\"adtplatz 6, A-1200 Wien, Austria
\and Wolfgang-Pauli-Institute c/o Faculty of Mathematics, University of Vienna, Oskar-Morgenstern-Platz 1, A-1090 Wien, Austria
\and Fakult\"at f\"ur Mathematik, Universit\"at Wien, Oskar-Morgenstern-Platz 1, 1090 Wien, Austria
}

\date{Received xxx / Accepted xxx, to be submitted to A\&A}
        
\abstract
   {Non-local models of stellar convection can account for mixing effects in regions adjacent to convectively unstable layers
     and for changes to the mean temperature structure caused by free, buoyancy driven convection. The physical 
     completeness of such models, however, depends on how third order correlations, which characterize the
     non-local transport processes, are expressed in terms of second order correlations and the stellar mean structure.}
   {Physical arguments and 3D hydrodynamical simulations were used to develop and test new closure relations for the skewness of
    the vertical velocity and temperature fields and third order cross-correlations to improve the predictive capabilities
    of non-local models of convection used in stellar astrophysics and in other disciplines such as meteorology.}
   {The structural form of the closure correlations was developed by a series of physical arguments and their accuracy was evaluated
     through self-consistency tests based on 3D hydrodynamical simulations for the Sun and a DA type white dwarf.}
   {The new closure relations derived for the skewness of vertical velocity and temperature fields
     provided improvements of up to an order of magnitude compared to previous models.
     This allows releasing the full potential of closure relations for the vertical velocity and temperature cross-correlations 
     previously proposed in meteorology as well as the construction of new, more reliable models for the third order
     moments of vertical velocity and temperature in non-local models of turbulent convection.}
   {The new models for the skewness and third order cross-correlations of vertical velocity and temperature permit the
    construction of non-local models of turbulent convection which remove, among others, several major short-comings of three 
    equation non-local convection models that are based on the downgradient approximation. Their application in stellar evolution 
    modelling will help to clarify whether this approach is a major step forward in the modelling of stellar convection when
    multidimensional hydrodynamical simulations are not affordable.}

\titlerunning{Modelling TOMs and skewness}
\authorrunning{Kupka}
\maketitle
\nolinenumbers

\section{Introduction}   \label{sec:intro}

The modelling of energy transport, the mixing of chemical elements, and the excitation and damping of waves 
due to free, buoyancy driven convection is a major challenge in stellar physics \citep{cox04b,kupka17b} and
also in the related disciplines of planetary sciences and geophysics (see \citealt{mironov09b} for a review
on numerical weather prediction). This also holds when neglecting the interactions of convection with rotation 
and with magnetic fields (reviewed in \citealt{brun17b}). For stellar evolution the modelling of convection 
limits the accuracy and precision to which pre-main sequence evolution tracks as well as stellar radii during 
the main sequence and later stages of stellar evolution can be determined. Its interaction with nucleosynthesis 
and the mixing of chemical elements throughout stellar evolution restrains the predictive capability of 
stellar models (for introductions see \citealt{Hansen2004} and \citealt{Kippenhahn2012}). The modelling
of convection is also important to helio- and asteroseismology, as convection drives and damps stellar 
pulsation and can be probed by seismology \citep{houdek15b,JCD21}. 

Mixing length theory (MLT) \citep{biermann32r,bv58b} has long remained the standard model for stellar 
convection \citep{cox04b}. Its long lasting success (cf.\  \citealt{Hansen2004} and \citealt{Salaris2005}) has been
accompanied by the discovery of an increasing number of its limitations. These include convective overshooting
and its impact on the main sequence turn-off in stellar clusters \citep{VandenBerg2006} as well as on
the HR diagram location of eclipsing binaries \citep{claret2019}. The structural surface effect discovered
in the analysis of the discrepancies between observed and predicted frequencies of solar p-modes is a term
specifically invented to describe a major failure of MLT based solar models (see \citealt{JCD02}, 
\citealt{houdek15b}, \citealt{JCD21}) and similar mismatches can be found for sound speed predictions at
the bottom of the solar convection zone \citep{christensen2011}. This has partially been remedied 
by either calibrating the mixing length itself to match the specific entropy predicted by
3D numerical simulations \citep{Spada21} or by replacing the predictions for the stellar surface layers
with averages from 3D hydrodynamical simulations altogether, a technique know as model patching
\citep{rosenthal99b,Mosumgaard2020}. This approach is limited to modelling stellar surface
layers, since for the deep stellar interior the thermal relaxation of 3D hydrodynamical simulations 
becomes unaffordable \citep{kupka17b}. 

A revisit to a long known alternative is hence worthwhile: non-local models of convection
which require the construction of differential equations for predicting statistical properties of
the dynamical variables describing convection, such as  the velocity $\vec{u}$ and temperature $T$.
Examples for them include the models of \citet{gough77b}, \citet{xiong78b}, \citet{stellingwerf82b}, 
\citet{kuhfuss1986}, \citet{xiong97b}, \citet{canuto92b,canuto97b,canuto09b,canuto11b}, and 
\citet{li12b}. In each of them the non-linearity of the advection operator in the Navier-Stokes equations 
leads to a closure problem: higher order moments have to be expressed in terms of lower order ones to 
derive a closed set of dynamical equations for the statistical properties of turbulent flows \citep{pope00b}. 

Non-local models of convection have often been used in 1D modelling of stellar pulsation but less frequently
in stellar evolution models. Models used in practice mostly add just one dynamical equation
to the equations describing the mean structure of the star (one equation models). Examples used 
in stellar pulsation and stellar evolution codes are the models of \citet{stellingwerf82b}, \citet{kuhfuss1986}, 
\citet{roxburgh89b}, and \citet{zahn91b}. A common additional dynamical variable is the turbulent kinetic 
energy. Its advection into or away from a local point is computed from a third order moment (TOM), the flux of 
kinetic energy $F_{\mathrm{kin}}$. This is accounted for by \citet{stellingwerf82b} and \citet{kuhfuss1986}: their 
models can also be used to model time dependent changes of the stellar structure due to large amplitude radial 
pulsations. The models of \citet{roxburgh89b} and \citet{zahn91b} neglect $F_{\mathrm{kin}}$ and assign all non-radiative 
transport to the convective enthalpy flux $F_{\mathrm{conv}}$ which restricts them to stellar evolution modelling.
In \citet{stellingwerf82b} and \citet{kuhfuss1986} $F_{\mathrm{kin}}$ is modelled as diffusion down a gradient.
The idea of assuming a downgradient model for all third order moments describing transport by advection 
was used by \citet{xiong78b,xiong85b,xiong86b} and independently by \citet{kuhfuss1986,kuhfuss1987}. 
It was also suggested by \citet{canuto98b} as the most simple approach to close their Reynolds stress model
of stellar convection.

Except for the most simple one-equation models of convection, which neglect $F_{\mathrm{kin}}$ 
altogether, non-local models of turbulent stellar convection require statistical (closure) approximations 
of TOMs. A new model for the computation of the TOMs describing non-local transport was hence developed 
with the aim to overcome some of the limitations of the popular downgradient approximation (DGA). It is based 
on ideas previously suggested in meteorology (\citealt{zilitinkevich99b}, \citealt{gryanik02b}, \citealt{gryanik05b}).
The latter required an external specification of the skewness of the fields of vertical velocity $w$ and 
temperature $\theta$, $S_{w}$ and $S_{\theta}$, respectively. To avoid this limitation new models for 
$S_{w}$ and $S_{\theta}$ were developed which allow the construction of a complete closure of the TOMs
$\overline{w^3}$, $\overline{w^2\theta}$, and  $\overline{w\theta^2}$ in terms of second order moments and 
mean structure variables without assuming the plain DGA. The model structure is motivated in 
Sect.~\ref{sect_construct} and its equations are derived in Sect.~\ref{sect_model}. Consistency tests of
the model based on 3D hydrodynamical simulations are presented in Sect.~\ref{sect_tests}.
A discussion and conclusions concerning the model equations are given in Sect.~\ref{sect_discussion}
and~Sect.~\ref{sect_conclusions}.

\section{Preparation of the model construction}    \label{sect_construct}

The derivation of the model equations relies on physical concepts revisited in the following.
They motivate the structure of the new models for the skewness of vertical velocity, $S_w$, 
and temperature, $S_{\theta}$. Formally, $S_w$ and  $S_{\theta}$ are defined as 
\begin{eqnarray}
S_w          & = & \overline{w^3} / \left(\overline{w^2}\right)^{3/2}, \\
S_{\theta} &  = & \overline{\theta^3} /  \left(\overline{\theta^2}\right)^{3/2}.
\end{eqnarray}

The plain DGA assumes each TOM to just depend on a single gradient. Almost all applications of 
non-local models of convection in astrophysics have been based on that assumption. In three-equation 
models, which consist of dynamical equations for $\overline{w^2}$, $\overline{w\theta}$, and 
$\overline{\theta^2}$, the DGA is defined by the following equations:
\begin{eqnarray}
    \overline{w^2\theta} & = & -D_1\, \partial{\overline{w\theta}}  / \partial  z,  \label{eq_w2t_DGA} \\
    \overline{w\theta^2} & = & -D_2\, \partial{\overline{\theta^2}} / \partial  z,  \label{eq_wt2_DGA} \\ 
    \overline{w^3}          & = & -D_3\,  \partial{\overline{w^2}}      / \partial  z,   \label{eq_w3_DGA}
\end{eqnarray}
where the turbulent diffusivities $D_i$ each depend on $\overline{w^2}$, a mixing length or dissipation
time scale, and other local quantities. Several, very similar variants of this model have been suggested
(see \citealt{xiong78b}, \citealt{kuhfuss1987}, or \citealt{canuto98b}).

\subsection{Mass flux models and statistical distribution functions}    \label{sect_existing}

\subsubsection{Coherent structures and their modelling}

Convection itself is not necessarily turbulent. A convective instability in a fluid stratified
by gravity (see \citealt{cox04b}, \citealt{Hansen2004}, or \citealt{Kippenhahn2012}) 
first of all leads to up- and downflows along the directions of the oppositely acting
forces of buoyancy and gravitation. Turbulence can be triggered by convection as a secondary 
effect such as friction at a solid surface as in the planetary boundary layer (PBL) of the atmosphere 
of the Earth (see \citealt{mironov09b} for further literature) or by shear forces between up- and downflows 
or horizontal shear stresses due to rotation. Each of these can give rise to a Kelvin-Helmholtz instability
\citep{Lesieur08b} which for sufficiently large length scales and/or velocities can let the flow become 
turbulent. For convection in non-degenerate stars and in white dwarfs this is essentially always the 
case (see \citealt{Muthsam10a} or \citealt{kupka17b}) and thus buoyancy becomes a source 
of turbulence (see \citealt{canuto98b}, \citealt{canuto09b}).

Convection can be driven also just in a narrow region where heat transport by conduction or radiation
is strongly inhibited, as in the hydrogen and helium partial ionization layers in the envelope of cool stars. 
Alternatively, it may be driven in an extended region, for example in the core region of massive stars
where local heat production due to CNO cycle hydrogen burning is large. The meteorological phenomena 
of dry convection above a heated terrestrial surface and moist convection in clouds (cf.\ 
\citealt{mironov09b}) are more closely related to the first case for which observations as well as 
3D numerical simulations reveal the formation of large scale coherent structures. These can be 
represented by a splitting into separate regions of up- and downflow. In meteorology this initiated
the development of the mass flux model of convection (\citealt{arakawa69b}, \citealt{arakawa74b}, 
see \citealt{mironov09b} for a review). Once a flow has been set into motion it may readily pass through 
locally stable regions simply by Newton's law of inertia (see \citealt{canuto09b}). Independently thereof, 
two-stream models had been suggested to better explain solar observations (see \citealt{Ulrich1970} 
and references therein) and more general two-component model stellar atmospheres had been developed 
in that period \citep{Nordlund1976}. These models were based on observational constraints 
and on imposing physical conservation laws. By contrast the MLT picture of a bubble moving along in an 
unstably stratified layer is tied to local linear stability analysis \citep{schwarzschild06r} and therefore cannot 
provide a reliable prediction of flow topology, the extent of turbulently mixed regions, or the level of turbulence 
in a flow. The implications of the fundamentally non-local nature of stellar convection were discussed in 
\citet{Spruit1997} (cf.\ \citealt{stein98b}) and inspired a generalization of MLT: it identifies surface cooling 
and the ``entropy rain'' it generates as main drivers of convection in stellar envelopes \citep{Brandenburg2016}.
The physical soundness of this concept has recently been corroborated by \citet{Kaepylae2025} who
demonstrated by means of 3D numerical simulations of compressible convection that non-uniform, localized surface 
cooling drives deeply penetrating plumes as suggested by the model of  \citet{Brandenburg2016}. Experimental
evidence for narrow downflow regions deeply penetrating the upper solar convection zone has recently been
suggested to result from interpreting data obtained by local helioseismology \citep{Hanson2024}.

\subsubsection{Filling factors, plume models, and skewness}

Separate averaging over up- and downflows can be parametrized by a filling factor $\sigma$,
the ratio of areas with upflow in a layer relative to its total area. This allows to construct TOMs
for non-local Reynolds stress models of convection (see \citealt{canuto98b} and for a generalization \citealt{canuto07b}):
\begin{eqnarray}
    \overline{w^2\theta} & = & S_w\, \left(\overline{w^2}^{1/2}\right)\, \overline{w\theta}  \label{eq_w2t_CD98} \\
    \overline{w\theta^2} & = & S_w\, \left(\overline{\theta^2}^{1/2}\right)\, \overline{w\theta}   \label{eq_wt2_CD98} \\ 
    \sigma & = & \frac{1}{2}  \left(1-S_w\,\left(4+S_w^2\right)^{-1/2}\right). \label{eq_sigma}
\end{eqnarray}
In this model, $\sigma$ is only a diagnostic quantity parametrized by $S_w$. \citet{canuto98b} 
suggested to use the full TOM equation for $\overline{w^3}$ for its computation. This required
to also model the TOMs $\overline{q^2\,w}$ and $\overline{q^2\,\theta}$ in the same framework
($\overline{q^2} = \overline{u^2} + \overline{v^2} + \overline{w^2}$ is the total specific turbulent kinetic energy,
the sum of contributions by horizontal and vertical velocities, $\overline{u^2} + \overline{v^2}$ and $\overline{w^2}$, 
respectively). The equations for these TOMs were closed in \citet{canuto98b} by 
the quasi-normal approximation (QNA) for fourth order moments with eddy damping.\footnote{
        The QNA of fourth order moments by \citet{millionshchikov41b} assumes that all even order 
        moments higher than second order are normally distributed. This allows closing the dynamical equations
        of the TOMs (cf.\ \citealt{canuto92b}). To ensure realizability the TOMs either have to be flux limited or eddy
        damping has to be applied to the pressure correlations in their dynamical equations (see \citealt{Lesieur08b}).
        The advanced non-local model of deep convection by \citet{canuto07b} also avoids the simple QNA.}
An equation for $\overline{w^3}$ was obtained which is a linear combination of the gradients for $\overline{w^2}$, $
\overline{q^2}$, and $\overline{w\theta}$ multiplied with highly non-linear front factors. It allowed the computation 
of $S_w$ in terms of second order moments and mean values. In spite of the algebraic form of 
(\ref{eq_w2t_CD98}) and (\ref{eq_wt2_CD98}), in that model $\overline{w^2\theta}$ and $\overline{w\theta^2}$ 
become proportional to a linear combination of several gradients of second order moments.

A physically more complete but also more complex alternative was suggested in \citet{canuto98b} as well: to use 
the stationary limit full solution for the dynamical equations of the TOMs $\overline{w^3}$, $\overline{\theta^3}$,
$\overline{w^2\theta}$, $\overline{w\theta^2}$, $\overline{q^2\,w}$, and $\overline{q^2\,\theta}$ closed by an 
eddy damped QNA which can be expressed as a linear combination of all the gradients of the different second 
order moments \citep{canuto93b}. Thus, both models suggested to compute the TOMs from linear 
combinations of gradients of second order moments multiplied with highly non-linear algebraic expressions.

Assuming the mass flux approximation, independently of \citet{canuto98b}, \citet{Abdella97} also arrived at 
Eq.~(\ref{eq_wt2_CD98})--(\ref{eq_sigma}) and suggested a simpler approximation to compute $S_w$. In 
the end they arrived at a linear combination of gradients of second order moments to compute the TOMs, too. 

As will be discussed below, Eq.~(\ref{eq_w2t_CD98}) is a much better model for $\overline{w^2\theta}$ than 
Eq.~(\ref{eq_w2t_DGA}) provided $S_w$ is not computed from a plain or generalized DGA model for 
$\overline{w^3}$. Eq.~(\ref{eq_wt2_CD98}) has an additional shortcoming. \citet{mironov99b} noted that contrary to 
Eq.~(\ref{eq_w2t_CD98}) it is not invariant under sign change of $w$ or $\theta$: for the product of  
$S_w\, \overline{w\theta}$, independently of the sign of $\theta$, $w$ can only be linked to non-negative contributions 
independently of its sign. This is how $\overline{w^2\theta}$ and its closure, Eq.~(\ref{eq_w2t_CD98}), 
behave, but it is not valid for $\overline{w\theta^2}$. And vice versa $\theta$ only leads to non-negative 
contributions in $\overline{w\theta^2}$ while this is not the case for $S_w\, \overline{w\theta}$ where a sign 
change of $\theta$ implies identical sign changes on both sides of Eq.~(\ref{eq_w2t_CD98}). To strengthen 
their argument \citet{mironov99b} demonstrated that the flux of temperature fluctuations, $\overline{w\theta^2}$, 
in 3D large eddy simulations (LES) data as well as in measurements of the convective PBL of the atmosphere 
of the Earth shows a negative sign above the convectively unstable layer, i.e.\ in its overshooting zone, while 
Eq.~(\ref{eq_wt2_CD98}) predicts the opposite. This happens in spite of a rather good agreement of this closure 
with the data throughout most of the interior of the convective zone. Their finding was independently confirmed 
in \citet{kupka07f} by means of 3D direct numerical simulations of fully compressible convection for 
both overshooting above and below convection zones. A similar problem was also found by 
\citet{kupka07d} for the DGA approximation of $\overline{w\theta^2}$, Eq.~(\ref{eq_wt2_DGA}).

\subsubsection{Probabilistic interpretation and two-scale mass flux approximation}   \label{sect_tsmf}

\citet{zilitinkevich99b} suggested to reinterpret the bimodal model used for the derivation of
Eq.~(\ref{eq_w2t_CD98}) in terms of probabilities. The splitting into up- and downflows depending 
on the local sign of vertical velocity is associated with the computation of the probability to find a particular 
value in a horizontal layer: subtracting the mean values of velocity and temperature from local values, 
assuming that basic conservation laws hold, and constructing the probability density function 
after averaging over areas of equal sign of $w$, they rederived the result for $\overline{w^2\theta}$ 
obtained by \citet{Abdella97} and \citet{canuto98b}, Eq.~(\ref{eq_w2t_CD98}). \citet{zilitinkevich99b} also pointed 
out that any good closure model ought to have the following properties: dimension, tensor rank and type, 
symmetries including sign changes, and realizability should be identical for both the quantity to be modelled 
and its closure approximation. \citet{mironov99b} used this probabilistic interpretation of \citet{zilitinkevich99b} 
to remedy the sign symmetry problem of Eq.~(\ref{eq_wt2_CD98}). They suggested to extend the bimodal model 
to a scenario with three different states and their associated probabilities and assigned a non-zero probability 
for cold updrafts to exist along hot updrafts and cold downdrafts. 
Accounting for this additional physical state they derived a new closure for $\overline{w\theta^2}$ :
\begin{equation}
    \overline{w\theta^2} \sim S_{\theta}\, \left(\overline{\theta^2}^{1/2}\right)\, \overline{w\theta}.  \label{eq_wt2_M99} 
\end{equation}
\citet{zilitinkevich99b} also suggested an extended advection plus diffusion closure for
Eq.~(\ref{eq_w2t_CD98}): 
\begin{equation}
\overline{w^2\theta} = a_1\,S_w\, \left(\overline{w^2}^{1/2}\right)\, \overline{w\theta}  
                                   -D_4\, \partial{\overline{w\theta}}  / \partial  z \label{eq_w2t_Z99}
\end{equation}
This suggestion\footnote{Originally, another additive term proportional to the superadiabatic gradient 
                          was suggested. Following \citet{mironov99b} and \citet{gryanik02b} it was neglected in the following.}
was based on the notion that both Eq.~(\ref{eq_w2t_CD98}) and Eq.~(\ref{eq_w2t_DGA}) 
cannot be reduced to each other. Eq.~(\ref{eq_w2t_CD98}) represents perfect mixing with no vertical gradients 
while the opposite case leads to strong gradients. These two cases are irreducible:
in a perfectly mixed flow there are no gradients and thus the DGA and the eddy damped QNA would
predict the TOMs to be zero while Eq.~(\ref{eq_w2t_CD98}) and Eq.~(\ref{eq_wt2_M99}) can 
remain non-zero: as for the entropy rain model of \citet{Brandenburg2016} and as discussed for 
the convective (enthalpy) flux in \citet{canuto09b} (Eq.~(37)--(38) therein) a convective heat flux may be 
generated in a thin layer and transported over extended distances without requiring a driving gradient in the region
through which the energy is transported. 

Such a flow state has also been called ballistically stirred by \citet{gryanik02b}. Their work
introduced the two-scale mass flux approximation (TSMF) which suggests separate averaging
for each of the different combinations of signs for $w$ and $\theta$, four in total. This means
to also account for hot downflows in addition to the states considered in \citet{mironov99b}.
For the TOMs they recovered the closures of \citet{zilitinkevich99b} and \citet{mironov99b}.
They additionally suggested to also add gradient diffusion to the closure for $\overline{w\theta^2}$,
\begin{equation}
    \overline{w\theta^2} = a_2\,S_{\theta}\, \left(\overline{\theta^2}^{1/2}\right)\, \overline{w\theta}
                                        -D_5^{\prime}\, \partial{\overline{\theta^2}} / \partial  z,  \label{eq_wt2_GH2002} 
\end{equation}
and, in agreement with \citet{zilitinkevich99b}, noted after comparisons with numerical simulations of 
the PBL of the Earth atmosphere that $a_1=1$ and $a_2=1$ yields sufficiently accurate results. 
\citet{gryanik02b} also suggested closures for many of the fourth order moments based on the 
two-scale mass flux approach and their approach was further generalized in \citet{gryanik05b} 
to horizontal velocities. Very good agreement for these models was found in comparisons of the model 
predictions with aircraft measurements of the PBL and numerical 
simulations\footnote{It is interesting to note that for the PBL the non-Gaussian contribution 
                                  can be accounted for through proper eddy damping \citep{cheng05b},
                                  an approach which in the end also computes the TOMs from linear combinations 
                                  of gradients of second order moments.}
in the original papers. The models were also successfully validated 
in oceanography \citep{losch04b} and with 3D hydrodynamical simulations of surface convection in the Sun 
and a main sequence K dwarf \citep{kupka07b}. Confirming previous authors \citet{cai18b} found the TSMF 
model to outperform the QNA when probing it with 3D numerical simulations of compressible convection 
by a factor of two. \citet{kupka07b} also noticed that the best of the two-scale mass flux 
closures of \citet{gryanik02b} was that one for $\overline{w\theta^2}$, Eq.~(\ref{eq_wt2_GH2002}), followed 
by that one for $\overline{w^2\theta}$, Eq.~(\ref{eq_w2t_Z99}), even without added gradient diffusion terms. They
noted ``a nearly perfect match'' when the latter were included. The excellent performance of Eq.~(\ref{eq_w2t_CD98}) 
was also shown in \citet{kupka17b} who analyzed 3D hydrodynamical simulations of solar surface convection 
with open vertical boundary conditions and simulations of a thin convection zone embedded in thick radiative 
zones in a DA white dwarf. They also demonstrated for Eq.~(\ref{eq_w2t_CD98}) that it is of 
secondary importance whether Favre averages are used in the computation:  
$\overline{\rho w^2 \theta}/\overline{\rho}$ and $\overline{w^2 \theta}$ agree mostly to better than 5\%.
Even at the superadiabatic peak their difference is less than 20\%.

In spite of their excellent performance these closures have not been used in calculations of stellar 
evolution, as they require accurate models for $S_w$ and $S_{\theta}$ to outperform DGA based 
closures. \citet{kupka07d} and \citet{kupka07f} concluded from their
direct numerical simulation of compressible convection that the DGA and the stationary limit 
full solution for the TOMs yield unsatisfactory results for $\overline{w^3}$ particularly for the 
case of a three pressure scale heights deep convection zone and poor results for 
$\overline{w^2\theta}$ and $\overline{w\theta^2}$ for all configurations of convective zones tested. 
The overall results for thin convective zones are acceptable when using the stationary limit 
full solution (\citealt{kupka99b}, \citealt{kupka02b}, \citealt{montgomery04b}) since 
$\overline{w^2\theta}$ and $\overline{w\theta^2}$ are less important in that 
case.\footnote{Its good agreement found in \citet{canuto94b} and \citet{cheng05b} for the PBL could 
         also be related to its much smaller depth $d \lesssim 0.125 H_p$ and the different boundary 
         conditions (solid bottom plate).} 
In contrast, with $S_w$ and $S_{\theta}$ taken from the numerical simulations, the performance of 
Eq.~(\ref{eq_w2t_Z99})--(\ref{eq_wt2_GH2002}) is an order of magnitude better than that of
Eq.~(\ref{eq_w2t_DGA})--(\ref{eq_wt2_DGA}) and for both the correct sign changes are recovered.

\subsection{The skewness of vertical velocity and temperature}    \label{sect_skewness}

A new model for $S_w$ and $S_{\theta}$ was hence constructed which contains an additive 
term that cannot be reduced to a linear combination of gradients of second order moments with 
multiplicative factors. This additive splitting into an algebraic term and a diffusion-type term was
motivated by extending several arguments provided in \citet{zilitinkevich99b} and \citet{canuto09b}.
Driving of convection within a narrow region as free convection occurs independently of the presence 
of shear stresses. It can be present even in physical systems with a laminar convective flow. Strong 
shear stresses in turn can lead to convective heat transport in systems with forced convection, 
with overshooting as a special case, and for situations without a clear correlation between up- and 
downflows as well as hot and cold flows. From a probabilistic viewpoint these two flow states can be
driven simultaneously. Due to their different origins, however, they can be considered as independent 
of each other which allows for irreducible states. The main hypothesis assumed in the new model is that 
this holds in general: interactions (including extreme events as discussed in \citealt{pratt17a}) are sufficiently 
rare and to first approximation they can  be neglected. The free convective (algebraic) and the forced (diffusion) 
contribution can thus be added, each of them equipped with leading multiplicative closure constants to adjust 
their contributions in the transition  (interpolation) regime, as has already been 
done\footnote{\citet{cheng05b} found the predictions for $\overline{w^4}$ of the model of 
      \citet{gryanik02b} to deviate from 3D simulation data near the top of the PBL. This agrees with the 
      latter stating that entrainment (overshooting) layers are outside the region of applicability for their 
      fourth order moment closure model.}
in the case of Eq.~(\ref{eq_w2t_Z99})--(\ref{eq_wt2_GH2002}).

A 3D LES of compressible convection by \citet{chan96b} provided first guidance for modelling 
$S_w$. They found the closure $\overline{w^3} \approx -1.45\left(\overline{w^2}\right)^{3/2}$ or 
$S_w \approx -1.45$ to work quite well for the lower four pressure scale heights of 
the convective zone. Deviations were reported to be large near the top of the convective zone and 
in the overshooting region. Eq.~(\ref{eq_w2t_CD98}) and~(\ref{eq_wt2_M99}) were not tested. 
Generally, they found pure algebraic closures to perform better inside the convection zone.
The opposite was concluded for diffusion-type closures. Remarkably, the closure
$\overline{w^3} \approx -1.9 \left(\overline{w^2}\right)^{1/2} \ell\, \partial{w^2}/\partial z$
with $\ell = H_p$, the local pressure scale height, worked well where the algebraic closure
performed poorly and vice versa: the DGA closure underestimated $S_w$
inside the convective zone and worked well near its boundary and 
in the overshooting zone. When adding both, the algebraic closures should dominate 
inside the convective zone and be negligible in the overshooting region. This is
indicative for
\begin{equation}
    \overline{w^3} = a_3  \left(\overline{w^2}\right)^{3/2} 
                              + d_6 \left(\overline{w^2}\right)^{1/2} \ell\, \partial{\overline{w^2}}/\partial z \label{eq_w3_CS96_added}                              
\end{equation}
with $a_3 \approx  -1.5$ and $d_6 < 0$ to be a first model for $S_w$. Eq.~(\ref{eq_w3_CS96_added}) can 
violate sign symmetry and its algebraic contribution remains constant independently of stratification. 
A solution to this limitation is suggested in Sect.~\ref{sect_model}. In \citet{kupka07b} $S_w$
is in the range of -1.4 to -1.5 within the first two pressure scale heights below the superadiabatic peak
for a solar surface simulation while it is closer to -1.25 for their K dwarf simulation which, however, is
even more strongly influenced by the closed vertical bottom boundary. With open vertical boundaries $S_w$ 
is usually between -1.5 to -1.8 in the interior of the solar convection zone (see \citealt{belkacem06b}
and  Appendix~\ref{app_plume}) and saturates at this value for several pressure scale heights. For narrow 
convection zones embedded in stably stratified layers no plateau was found as the overshooting layers 
directly interact with the entire convection zone. Even in that case a mean value around -1.5
appears to be realistic (cf.\ \citealt{kupka17b}). 

Numerical 2D and 3D simulations of convection in the deep stellar interior \citep{Dethero2024}
and theoretical considerations motivated by 3D direct numerical simulations of convection
\citep{Yokoi2022,Yokoi2023} indicate that both spatial and temporal correlations influence non-local
transport and the filling factor $\sigma$ and the skewness $S_w$. No simple predictive model for
the computation of $S_w$, however, could yet be derived and a more complete model would likely
introduce additional, dynamical equations for $\overline{w^3}$ and further quantities, which is
beyond the scope of the present work.

Comparing computations of $S_{\theta}$ from 3D RHD in \citet{belkacem06b}, \citet{kupka07b}, and 
\citet{kupka18b} this quantity is found to behave similar to $S_w$ inside convective
zones, where $w$ and $\theta$ are well correlated, as follows from 
\begin{equation}
  C_{w\theta} = \frac{\overline{w\theta}}{\left(\overline{w^2}\overline{\theta^2} \right)^{1/2}}    \label{eq_Cwt}   
\end{equation}
having positive values in the range of 0.7 to 0.85 in those layers (see also \citealt{chan96b}).
Thus, $S_{\theta} < 0$ inside stellar convection zones driven by cooling at the surface and 
$|S_{\theta}| \approx 1.4\, |S_w|$ (cf.\ \citealt{belkacem06b} and \citealt{kupka07b}).
This becomes more intricate for thin convective zones \citep{kupka18b}. To keep this first
model simple, $S_{\theta}$ was constructed on the basis of $S_w$ such that dimension, tensor rank, 
sign symmetry, and realizability are preserved, as reported in Sect.~\ref{sect_model}.

\section{Derivation of the model equations}    \label{sect_model}
\subsection{A model for $S_{\theta}$ to close $\overline{w\theta^2}$ in terms of $S_w$ or $\overline{w^3}$}    \label{sect_Stheta}

A simple closure for $\overline{\theta^3}$ in terms of $\overline{w^3}$ can be argued 
from dimensional constraints and sign symmetries: $\overline{w^2\theta}$ is invariant under
$w \rightarrow -w$ and changes sign if $\theta \rightarrow -\theta$. The relation
\begin{equation}   
   \overline{\theta^3} \sim  \overline{w^2\theta}  \, \frac{\overline{\theta^2}}{\overline{w^2}}   \label{eq_prop_t3}
\end{equation}
fulfils both and inserting Eq.~(\ref{eq_w2t_CD98}) into $\overline{w^2\theta}$ in Eq.~(\ref{eq_prop_t3}) results in
\begin{equation}
   \overline{\theta^3} \sim \overline{w \theta}  \,  \frac{\overline{w^3}}{\left(\overline{w^2}\right)^2}  \, \overline{\theta^2}.  \label{eq_t3}
\end{equation}
The contraction of the vectors $\overline{w \theta}$ and $\overline{w^3}$ in Eq.~(\ref{eq_t3})
ensures that $\overline{\theta^3}$ and thus $S_{\theta}$ are scalars.

The realizability of this model for $S_{\theta}$ is more obvious, if it is rederived by analyzing 
sign changes of $S_{\theta}$,  $S_w$, and $\overline{w\theta}$ as well as the convective and
kinetic energy fluxes, $F_{\rm conv}$ and $F_{\rm kin}$, in a 3D radiation hydrodynamical (RHD) 
simulation of a DA white dwarf with $T_{\rm  eff} =  11800\,\mathrm{K}$ from \citet{kupka18b}. Horizontal
averages of the simulation data over an extended time series reveal a sign change of $F_{\rm conv}$ 
at simulation box depths of $\approx 0.75\,\mathrm{km}$ and $\approx 2.8\,\mathrm{km}$ (with 
$F_{\rm conv} > 0$ inside that range, see their Fig.~3, the convectively unstable zone is located within 
$1\,\mathrm{km} \dots  2\,\mathrm{km}$, with vertical distances measured from the top of the simulation 
box). In the same data set the cross-correlation $\overline{w \theta}$ and the auto-correlation function 
$C_{w\theta} = \overline{w \theta} / (\overline{w^2}\,\overline{\theta^2})^{1/2}$ have a sign change also at 
$\approx 2.8\,\mathrm{km}$ and almost a sign change\footnote{Compressibility, non-local radiative cooling, and the sensitivity of 
               $C_{w\theta}$ to boundary conditions could explain that difference. Moreover, the non-radiative fluxes
               up there are small, with $|F_{\rm conv}| \lesssim 10^{-3}\,F_{\rm total}$.} 
($C_{w\theta} < 0.03$) at $\approx 0.7\,\mathrm{km}$.
$S_w$ has a negative sign between $\approx 0.2\,\mathrm{km}$ and $\approx 5.1\,\mathrm{km}$ as has $F_{\rm kin}$
(see their Fig.~11 and Fig.~3). Outside that region $|F_{\rm kin}| \lesssim 10^{-5}\,F_{\rm total}$. Another observation 
is that $S_{\theta} < 0$ between $\approx 1\,\mathrm{km}$ and $\approx 2.8\,\mathrm{km}$ (see their Fig.~11), 
i.e.\  roughly where  $S_w < 0$ and $C_{w\theta} > 0$ and thus $S_w C_{w\theta}  <  0$, too. Below 
$\approx 2.8\,\mathrm{km}$, $S_{\theta} >  0$ which due to the sign change of $C_{w\theta}$ agrees with 
$S_w C_{w\theta}  >  0$. Above the convective zone the relation between $S_{\theta}$ and $S_w C_{w\theta}$ 
is more complex, but this zone is subject to strong radiative cooling and non-radiative flux contributions
are very small. From this analysis, the relation
\begin{equation}
   S_{\theta} = a_4 \, C_{w\theta}\,  S_w    \label{St_closure}
\end{equation}
with $1 \lesssim a_4 \lesssim 2$ appears to be a useful first approximation in regions where non-radiative
energy flux contributions are larger in magnitude than 0.1\% of the total flux. Both sides of Eq.~(\ref{St_closure})
are dimensionless, feature sign symmetry (invariant under $w \rightarrow -w$ and changing sign with $\theta \rightarrow -\theta$),
the contraction of $C_{w\theta}\, S_w$ is a scalar as is $S_{\theta}$, and if $S_w$ is realizable, so is $S_{\theta}$,
since  $|C_{w\theta}| \leqslant 1$ by the Cauchy-Schwarz inequality and realizability can always be ensured by
choosing $a_4$ not too large. If Eq.~(\ref{eq_t3}) is divided by $(\overline{\theta^2})^{3/2}$ and $a_4$ 
is inserted on its right-hand side to replace similarity with equality, it is easy to see that Eq.~(\ref{eq_t3}) 
is equivalent to  Eq.~(\ref{St_closure}). If Eq.~(\ref{St_closure}) is used in Eq.~(\ref{eq_wt2_GH2002}) to eliminate 
$S_{\theta}$ and for generality $a_2\,a_4$ is replaced by $a_5$,  a new closure for $\overline{w\theta^2}$ is obtained 
which no longer depends explicitly on $\overline{\theta^3}$, but respects sign symmetries, contrary to Eq.~(\ref{eq_wt2_CD98}).
Its algebraic term reads
\begin{equation}
  \overline{w\theta^2} = a_5\, \left(\overline{w\theta}\right)^2  \, \overline{w^3}\, \left(\overline{w^2}\right)^{-2}
                                   = a_5\, S_w\,  \left(\overline{w\theta}\right)^2\, \left(\overline{w^2}\right)^{-1/2}  \label{eq_wt2_new}
\end{equation}
and it depends only on $\overline{w^3}$, $\overline{w^2}$, and $\overline{w\theta}$, as Eq.~(\ref{eq_w2t_CD98}) 
does for $\overline{w^2\theta}$. 

Following the arguments by \citet{zilitinkevich99b}, \citet{mironov99b}, and \citet{gryanik02b}, and
the discussion provided in Sect.~\ref{sect_skewness}, a gradient diffusion term should be added to
Eq.~(\ref{eq_wt2_new}) ($D_5$ can differ from $D_5^{\prime}$ in Eq.~(\ref{eq_wt2_GH2002})) to obtain
\begin{equation}
    \overline{w\theta^2} = a_5\,  \left(\overline{w\theta}\right)^2  \, \frac{\overline{w^3}}{\left(\overline{w^2}\right)^2}
                                        -D_5\, \partial{\overline{\theta^2}} / \partial  z,  \label{eq_wt2_with_grad} 
\end{equation}
which contrary to Eq.~(\ref{eq_wt2_GH2002}) no longer depends on $\overline{\theta^3}$.

\subsection{A model for $S_w$ and $\overline{w^3}$ to close $\overline{w^2\theta}$ and $\overline{w\theta^2}$}     \label{sect_Sw}

To close $S_w$, Eq.~(\ref{eq_w3_CS96_added}) for $\overline{w^3}$ is first divided by $(\overline{w^2})^{3/2}$ to obtain
\begin{equation}
    S_w = a_3 + d_6  \ell\, \overline{w^2}^{\,-1} \partial{\overline{w^2}}  / \partial z. \label{eq_Sw_CS96_added}                              
\end{equation}
The value for $a_3$ is not universal. In principle, it is a function constrained by the conservation of mass, 
momentum, and energy as well as other physical requirements. But deriving a prognostic equation for $a_3$ 
(or $S_w$) would mean to add a new, non-local and highly non-linear (partial) differential equation to 
the convection model. Given the status quo in convection modelling in stellar structure and stellar evolution 
theory such a complex model could be seen as unviable, since for the codes developed for this purpose 
numerical robustness is a requirement which is equally important as is physical reliability. A more simple approach 
which still provides an order of magnitude improvement over Eq.~(\ref{eq_w3_DGA}) was hence considered.
Accounting for a possible dependency on the Brunt-V\"ais\"al\"a frequency $\tilde{N}$ 
(see Eq.~(\ref{eq_epsilon}) below), Eq.~(\ref{eq_Sw_CS96_added}) is generalized to
\begin{equation}
    S_w = a_6\,f(\tilde{N},\ell,\overline{w^2},\overline{q^2}) 
               + d_6  \ell\, \overline{w^2}^{\,-1} \partial{\overline{w^2}}  / \partial z, \label{eq_w3_formal}
\end{equation}
where $a_6$ should be chosen in the range of $-1.25$ to $-1.8$, if the convective flow is driven by cooling at 
an upper boundary, in agreement with results from the literature (see Sect.~\ref{sect_skewness}). 
It could be chosen in the range $0 < a_6 \lesssim 0.5$, if the flow is driven by heating by a bottom boundary layer 
(as expected from the data of  \citealt{mironov99b} and \citealt{gryanik02b}). It is set to 0 if neither 
applies. This does not resolve cases where a transition between these states occurs as a function of
location or time. For convective envelopes the first case applies. For convective cores the third case 
is more appropriate, but this requires further study with 3D numerical simulations. If both types 
of zones couple, the transition between both states has to be modelled.

In the following, only the case of convective driving by cooling at an upper boundary is considered.
The most important correction to the non-diffusive part of the model is the introduction of 
a function $f(\tilde{N},\ell,\overline{w^2},\overline{q^2})$ such that $f \rightarrow 1$ where 
$\tilde{N}=0$ and $f \rightarrow 0$ far away from the convective zone, just as the dissipation 
rate length scale in the same region \citep{Kupka2022}. Due to the radial dependence of $f$ 
and due to the gradient diffusion term the quantity $S_w$ is not just constant with depth.

In \citet{Kupka2022} the dynamical equation for the dissipation rate of turbulent kinetic energy
$\epsilon$, suggested in \citet{canuto98b}, was used to account for how waves excited in
stable stratification damp convective overshooting. This equation can also guide the modelling of $f$,
since the plumes generated by free convection also excite the waves in stably stratified layers that 
draw off their kinetic energy.  With $K=\overline{q^2}/2$ it reads
\begin{eqnarray}  \label{eq_epsilon}
   \partial_t \epsilon + D_{\rm f}(\epsilon) & = &
     c_1 \epsilon K^{-1} g \alpha_{\rm v} \overline{w\theta} - c_2 \epsilon^2 K^{-1}
       + c_3 \epsilon \tilde{N} \nonumber \\
   \mathrm{with} \quad \tilde{N} & \equiv & \sqrt{g \alpha_{\rm v} |\beta|} \quad = \quad \tau_{\rm b}^{-1},
\end{eqnarray}
for small Prandtl numbers and $\beta=-((\partial T/\partial z)-(\partial T/\partial z)_{\rm ad})$ is the superadiabatic gradient,
$g$ the local gravitational acceleration, $\alpha_{\rm v}$ the volume expansion coefficient, and
$\tilde{N}$ the Brunt-V\"ais\"al\"a frequency with its reciprocal, the buoyancy time scale 
$\tau_{\rm b}$. The values $c_1=1.44$ and $c_2=1.92$ are roughly in the middle of the typical range of 
values suggested in \citet{tennekes72b} and \citet{hanjalic76b}. A value of $c_3=0.3$ was suggested 
in \citet{canuto98b}. Instead of considering the DGA for $D_{\rm f}(\epsilon)$, the relations
\begin{equation}    \label{eq_diffeps}
   D_{\rm f}(\epsilon) \equiv \partial_z(\overline{\epsilon w})\qquad
   \mathrm{and}  \qquad \overline{\epsilon w} =  \frac{3}{2}\,\frac{\overline{q^2 w}}{2} \,\frac{1}{\tau}
\end{equation}
from \citet{canuto92b,canuto09b} are preferred in the following. They performed very well when tested
with 3D numerical simulations of convection in \citet{kupka07e} who also found that the DGA failed
to correctly predict $\overline{\epsilon w}$. 

In the following, contrary to \citet{Kupka2022} (see their Sect.~3.5 and 3.6) a slightly more accurate 
approximation was made for the stationary limit of Eq.~(\ref{eq_epsilon}) by considering Eq.~(\ref{eq_diffeps}) 
and approximate it as $\partial_z(\overline{\epsilon w}) = (c_5/\ell)\, (3/4)\,  (\overline{q^2 w}/\tau)$ to arrive at 
\begin{equation}   \label{eq_eps_alpha_NBV}
  \frac{3}{4}  \frac{\overline{q^2 w}}{\tau}  \frac{c_5}{\ell} = 2  c_1  g  \alpha_{\rm v} \overline{w\theta} / \tau - 2\, c_2 \epsilon / \tau + c_3 \epsilon \tilde{N}
\end{equation}
instead of approximating the non-local contribution by $D_f(\epsilon) \approx -\alpha_{\epsilon} \epsilon / \tau$.  
In the stably stratified region, once also $F_{\rm conv} < 0$ (plume dominated region), the buoyancy term
$2\, c_1 g \alpha_{\rm v} \overline{w\theta} / \tau$ becomes small (cf.\ Fig.~8 in \citealt{Kupka2022} and the discussion 
in their Sect.~3.6 of the results of \citealt{kupka02b} and \citealt{montgomery04b}). Thus, it was neglected or considered to be
absorbed into $- 2\, c_2 \epsilon / \tau$, as in \citet{Kupka2022}. Recalling from \citet{canuto98b} that the dissipation rate time
scale $\tau$ is related to the dissipation rate length scale $\ell$ and $\epsilon$ via $\tau  =  2\,K/\epsilon$,
$\epsilon = c_{\epsilon}\,K^{3/2} /\ell$, $\tau = 2\,\ell / (c_{\epsilon}\,K^{1/2})$, whence $\ell =  K^{1/2}\,\tau\,c_{\epsilon}/2$,
$c_{\epsilon}  \approx 0.8$, and after division by $c_2\,\epsilon^2/K$, Eq.~(\ref{eq_eps_alpha_NBV}) can be rewritten  into 
\begin{equation}   \label{eq_eps_q2w}
  \frac{3}{4}\, \overline{q^2 w}\,  \frac{c_5}{\ell\,\tau}  \frac{K}{c_2\,\epsilon^2} =  -1 + \frac{c_3}{2\,c_2} \frac{\tau}{\tau_{\rm b}}.
\end{equation}
Ignoring velocity asymmetry for the turbulent kinetic energy, $K = \overline{q^2}/2 \approx (3/2)\, \overline{w^2}$,
whence also $\overline{q^2 w} \approx 3 \overline{w^3}$. This step tacitly assumes the skewness of the
horizontal velocity components to equal that of vertical velocity. In a straightforward generalization one could instead
model the skewness of the total velocity field $\overline{q^2 w}/K^{3/2}$ and derive $S_w$ from a model 
of non-isotropic flow. 
That step is not necessary for 3-equation non-local models of convection such as that one of 
\citet{kuhfuss1987} (see \citealt{Kupka2022}) which assume isotropy of the velocity field and hence
it was not further considered in the following.
With these approximations Eq.~(\ref{eq_eps_q2w}) was rewritten in terms of $ \overline{w^3}$ into
\begin{equation}   \label{eq_eps_w3}
  \frac{3}{8}\, \frac{c_5}{c_2}\, \frac{\overline{w^3}}{\overline{w^2} \ell\,\tau^{-1}} =  -1 + \frac{c_3}{2\,c_2} \frac{\tau}{\tau_{\rm b}}.
\end{equation}
Introducing $S_w$ instead of $\overline{w^3}$ and rearranging Eq.~(\ref{eq_eps_w3}) yields
\begin{equation}   \label{eq_tau_Sw}
   \tau =   \frac{2\,c_2}{c_3}\,\left(1 + \sqrt{\frac{3}{8}} \, \frac{c_5}{c_2\,c_{\epsilon}}\, S_w\right)\,\tau_{\rm b}.
\end{equation}
With $c_4 = c_3  /  (c_2\,c_{\epsilon)})  \approx 0.2$ (see \citealt{Kupka2022}) and setting 
\begin{equation}   \label{eq_a6}
   a_6^{\prime} = -\frac{c_2\,c_{\epsilon}}{c_5}\,\sqrt{\frac{8}{3}}
\end{equation}
and recalling that $\overline{q^2}/2 = K \approx (3/2)\, \overline{w^2}$ and $a_6 \approx  a_6^{\prime}$, the result
\begin{equation}   \label{eq_Sw_new_no_DGA}
   S_w = a_6 \left(1 - c_4\,\ell\,K^{-1/2}\,\tau_{\rm b}^{-1}\right) =  a_6\,f(\tilde{N},\ell,\overline{w^2},\overline{q^2}) 
\end{equation}
was obtained. It is important to note that the sign of $c_5$ in Eq.~(\ref{eq_eps_alpha_NBV}) 
is not constrained to be positive, because the term it appears in approximates the divergence of
a flux, which can act as a source or a sink. For stellar envelope convection, $a_6$ can be taken
from the range of $-1.25$ to $-1.8$, as discussed, and $f(\tilde{N},\ell,\overline{w^2},\overline{q^2})
= f(\tau_{\rm b},\ell,K) = 1 - c_4\,\ell\,K^{-1/2}\,\tau_{\rm b}^{-1}$. Moreover, at the 
transition to unstable stratification, $\tilde{N} = 0$,  whence $S_w = a_6$ at that point. 

\begin{figure}[t]
\centering
\includegraphics[width=0.45\textwidth]{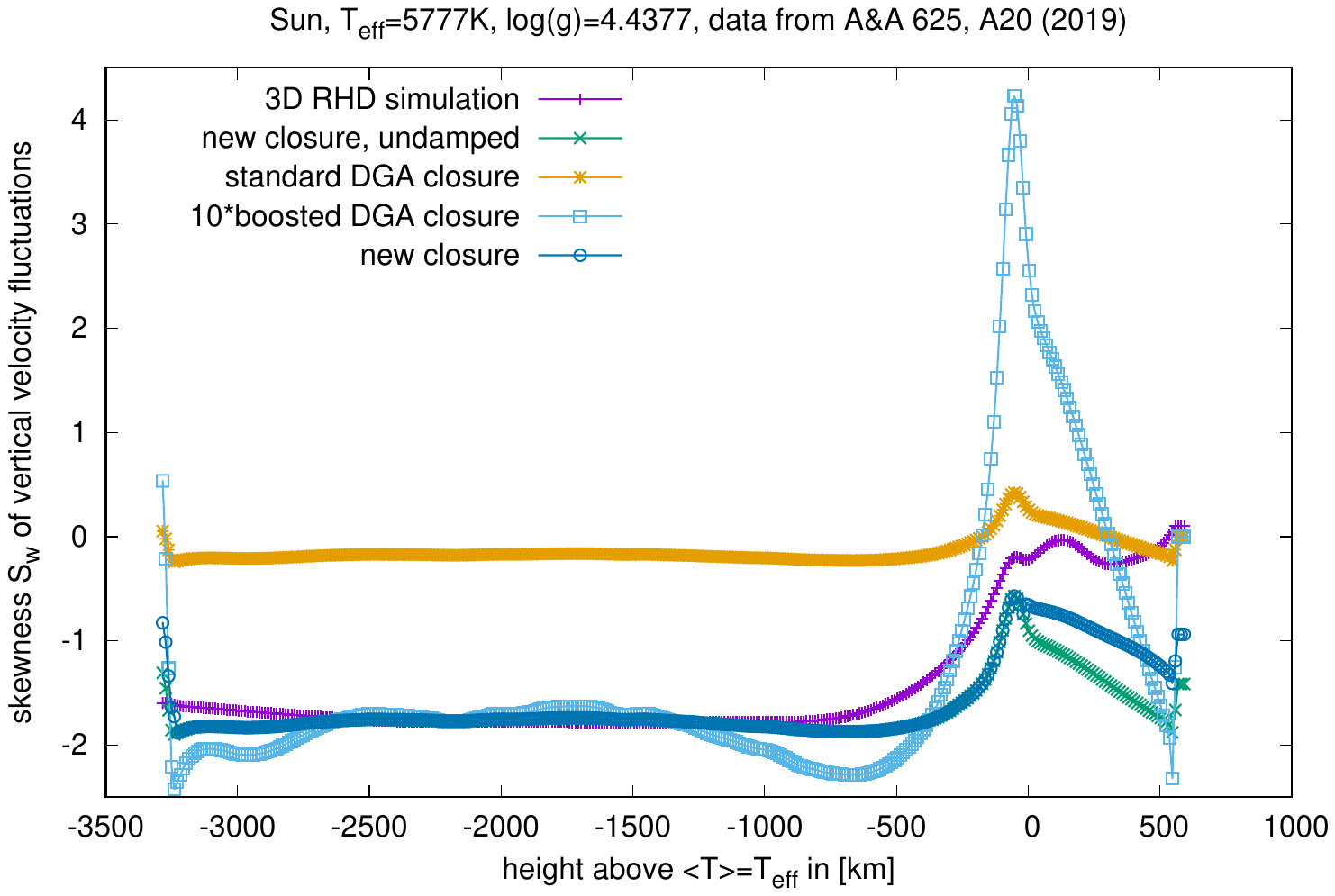}
\includegraphics[width=0.45\textwidth]{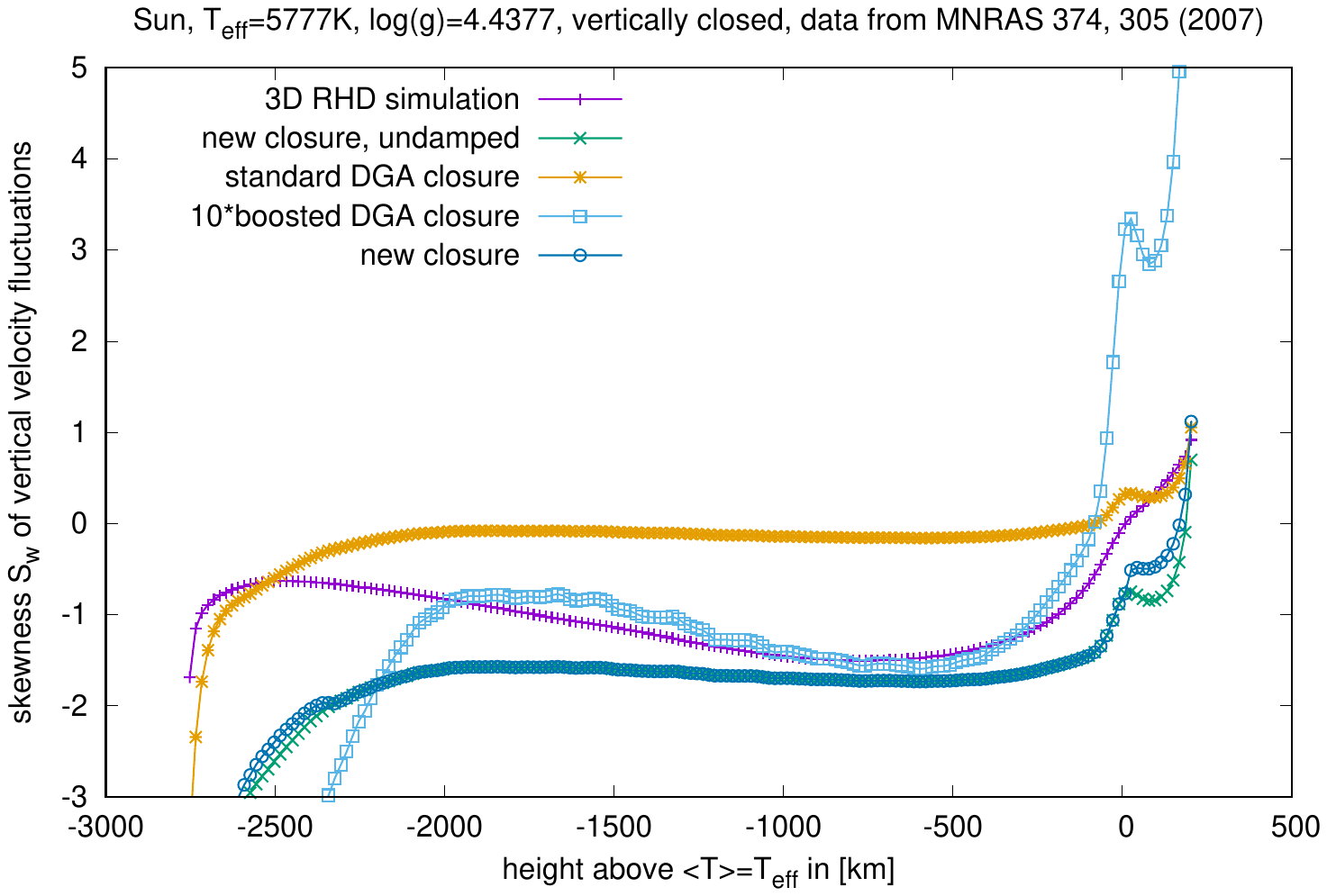}
\caption{Skewness of vertical velocity $S_w$ in 3D RHD simulations of solar granulation. Upper panel:
              open vertical boundary conditions at top and bottom (using the code of \citealt{Muthsam10a}).
              Lower panel: closed vertical boundary conditions (using the code of \citealt{robinson03b}).
              Data directly computed from the 3D RHD numerical simulations (purple line with crosses)
              are compared with the new TOM model and the DGA. For the new TOM model the green line 
              with x-shaped points shows the no damping case ($c_6=0$). The full, new model
              ($c_6=0.1$) is indicated by a dark blue line with circles. The light brown line with asterisks
              denotes the DGA ($d_3=-0.1$). The light blue line with squares shows the DGA with 
              ten times larger turbulent diffusivity ($d_3=-1$).
\label{Fig1}
}
\end{figure}

\begin{figure}[t]
\centering
\includegraphics[width=0.45\textwidth]{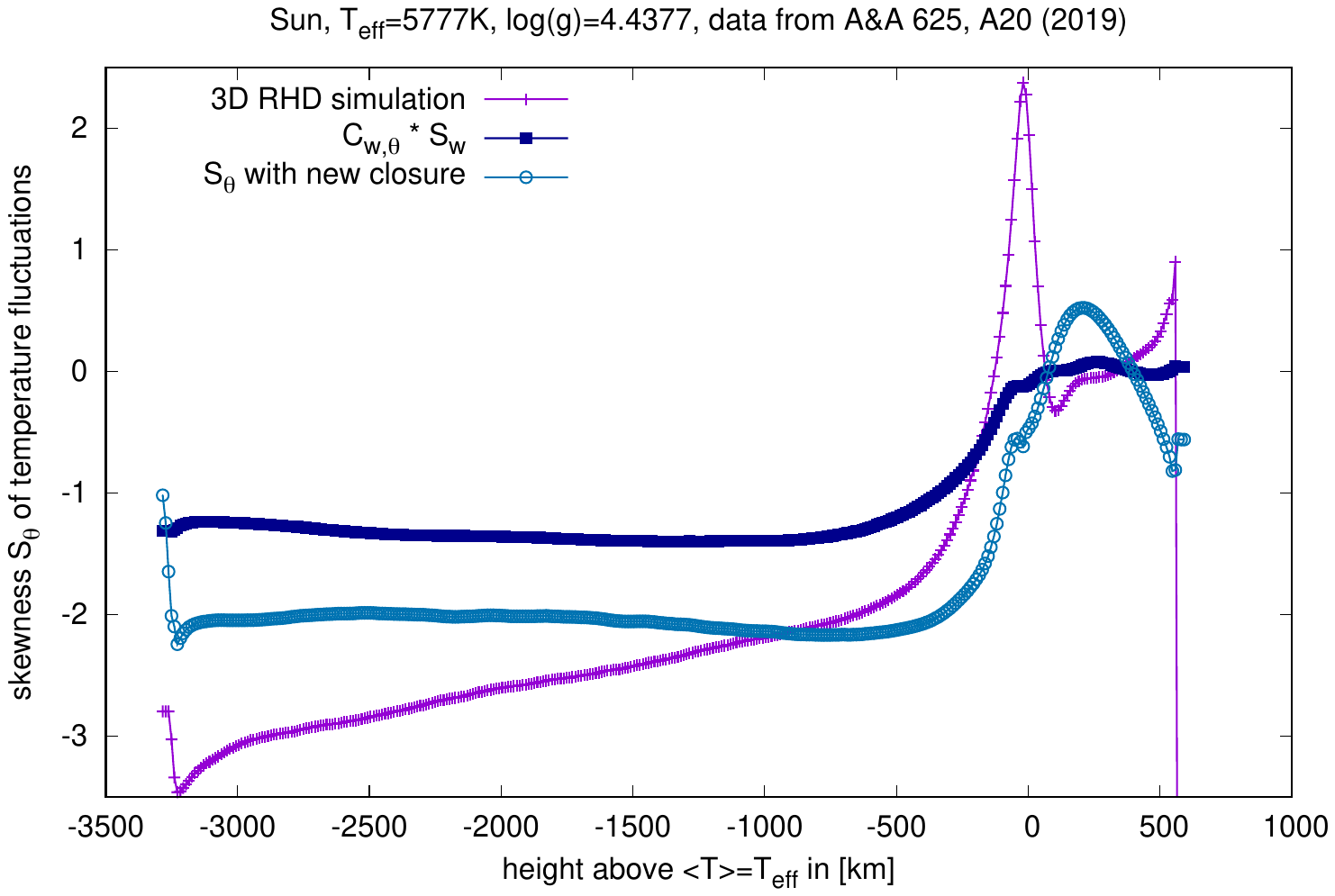}
\includegraphics[width=0.45\textwidth]{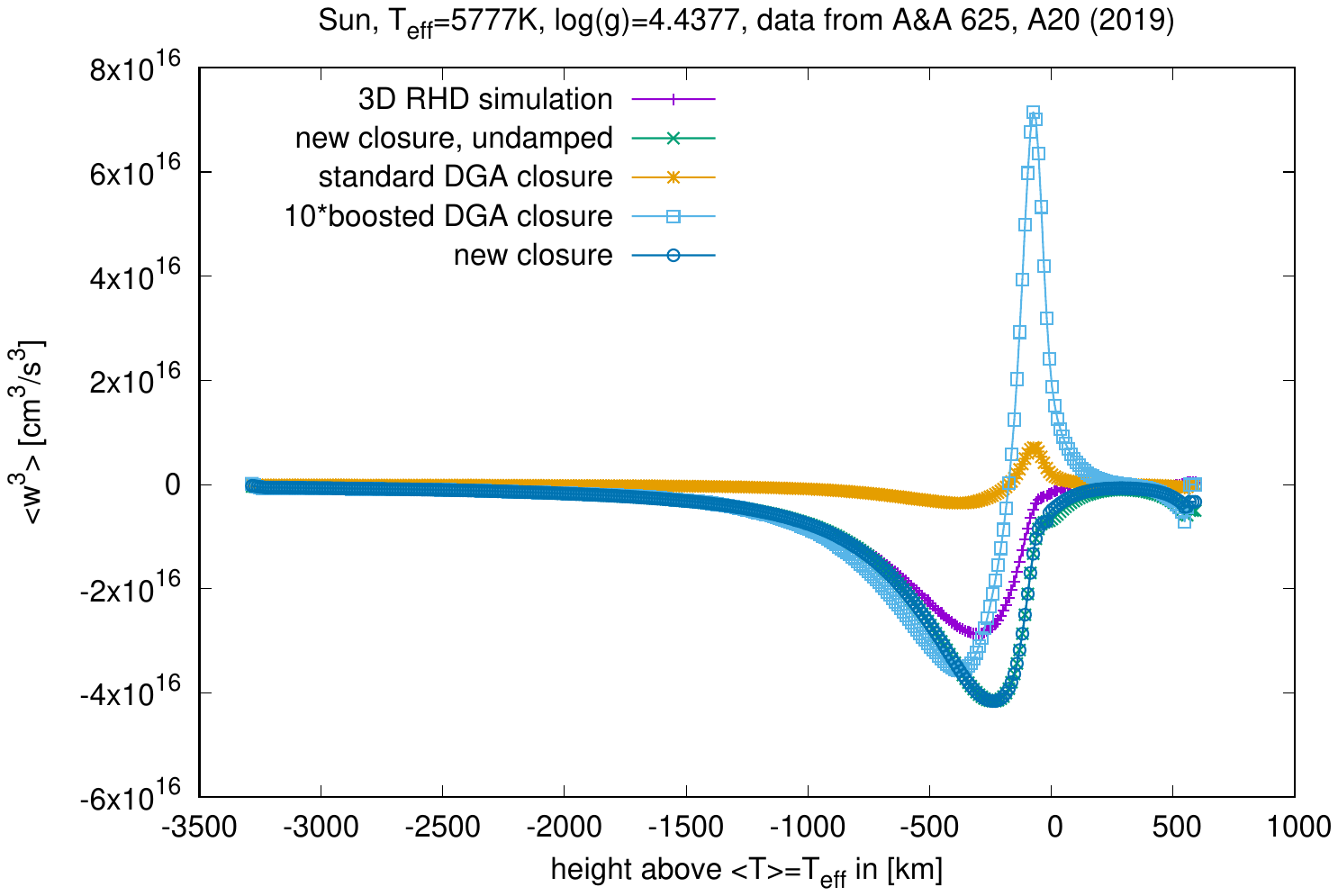}
\caption{Skewness of temperature $S_{\theta}$ (upper panel) and the TOM $\overline{w^3}$ (lower panel).
              The group of models displayed is the same as in the upper panel of Fig.~\ref{Fig1}. 
              Data were taken from the 3D RHD simulation with open vertical  boundary conditions as in the upper panel of Fig.~\ref{Fig1}.
              The same colour scheme was used as in Fig.~\ref{Fig1} except for the product of the cross-correlation $C_{w,\theta}$ 
              with $S_w$ (dark blue with filled squares as points). 
\label{Fig2}
}
\end{figure}

\begin{figure*}[t]
\centering
\includegraphics[width=0.45\textwidth]{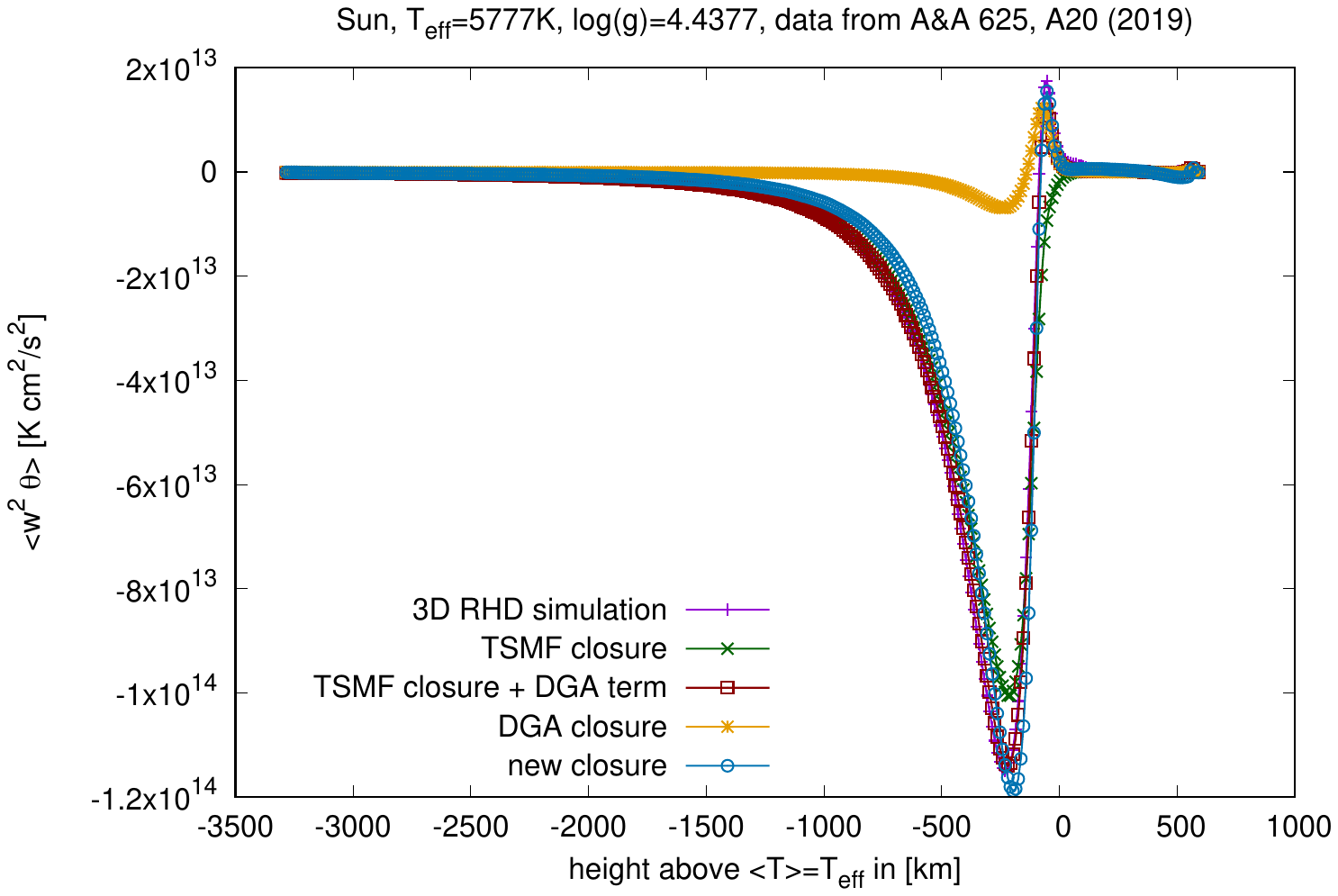}
\includegraphics[width=0.45\textwidth]{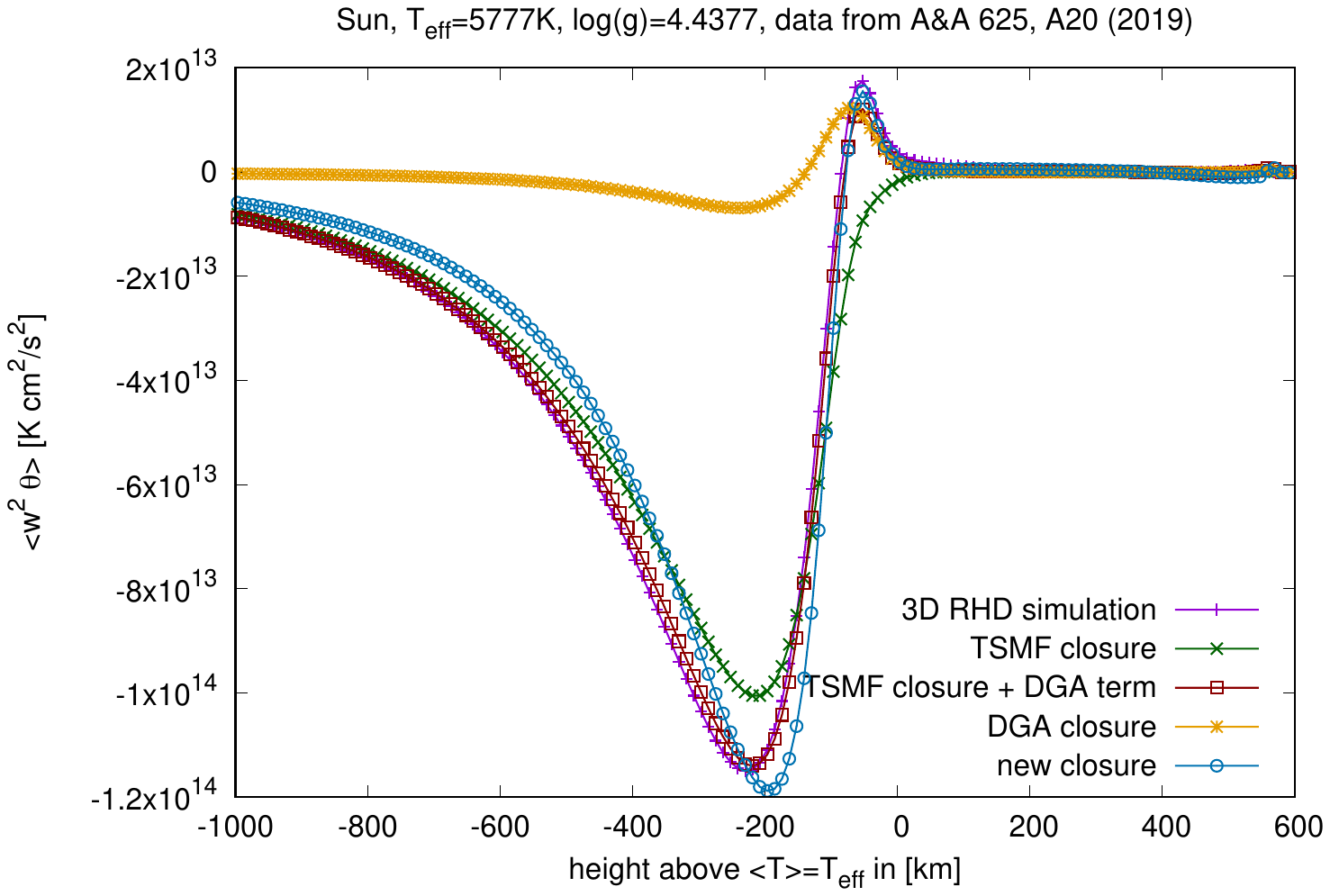}
\includegraphics[width=0.45\textwidth]{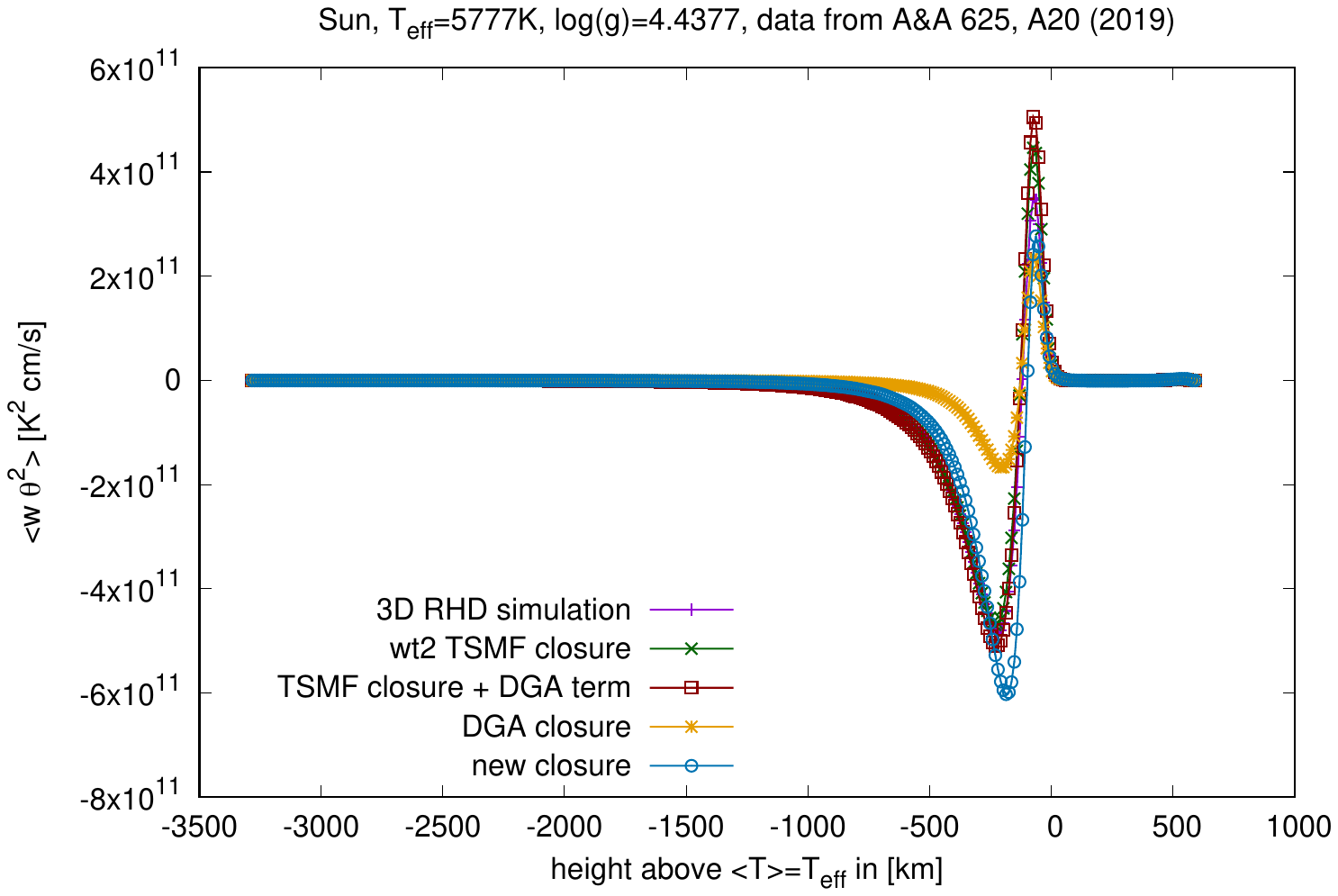}
\includegraphics[width=0.45\textwidth]{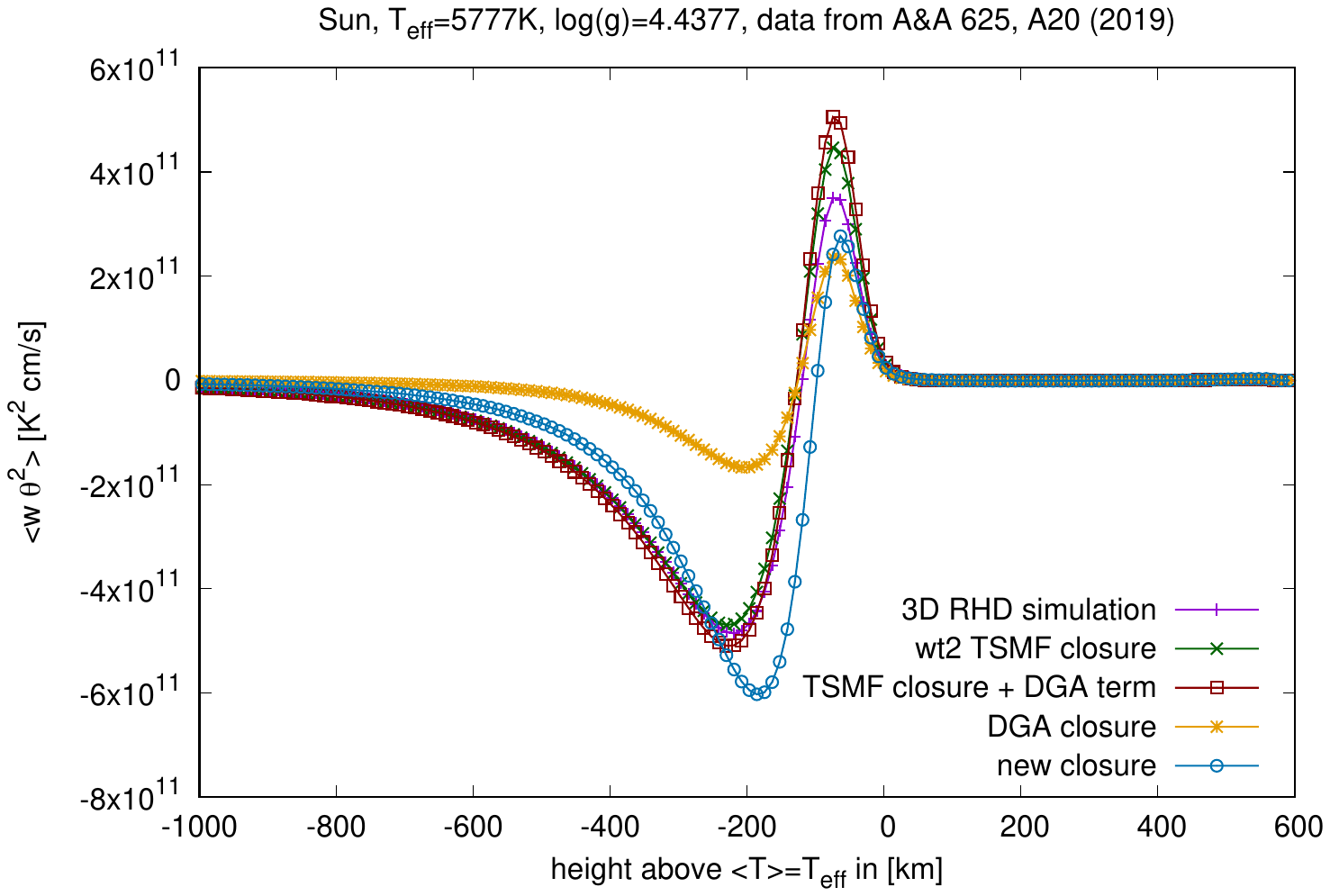}
\caption{Comparisons for the TOMs $\overline{w^2\theta}$ (upper row of panels) and $\overline{w\theta^2}$ (lower row of panels).
              Simulation data and models are the same as those studied in Fig.~\ref{Fig1} (upper panel) and Fig.~\ref{Fig2} (lower panel) 
              for $S_w$ and $\overline{w^3}$, respectively: each panel displays results for the solar surface 3D RHD simulation with 
              open vertical boundary conditions. The full vertical range is displayed in the left column of panels. The right column 
              of panels zooms into the upper 40\% of the simulation domain. The same colour coding was used as for Fig.~\ref{Fig1}. 
              In addition, the results obtained for the TSMF closure with DGA (dark red line with asterisks) and without (dark green line
              with crosses) are shown. For both variants of the TSMF model the quantity $S_w$ was computed directly from 3D RHD data,
              as explained in the main text.
\label{Fig3}
}
\end{figure*}

If the model is used in a wide parameter range, $c_4\,\ell\,K^{-1/2}\,\tau_{\rm b}^{-1} \approx 1$ 
may not always hold in the limit $\tilde{N} \gg 0$ in an overshooting layer far away from the convective 
zone. For instance, this criterion is roughly satisfied for the solar granulation simulations discussed in 
Sect.~\ref{sect_tests}, but it fails in the case of overshooting above and below a thin envelope convection zone of
a DA white dwarf discussed there as well, because $\tau \gg \tau_{\rm b}$. To ensure that $f \rightarrow 0$ for large 
$\tilde{N}$ a harmonic limiter was introduced into the model which forces $f$ to remain within $[0,1]$ for 
$c_4\,\ell\,K^{-1/2}\,\tau_{\rm b}^{-1} > 0$ and guarantees $f \rightarrow 1$ for $\tilde{N} \rightarrow 0$ and 
$f \rightarrow 0$ for $\tilde{N} \rightarrow \infty$. As the harmonic limiter overestimates $f$
where $\tilde{N}  \rightarrow 0$, $c_4$ should be replaced by $c_6$ with $c_4/2 \lesssim c_6 \lesssim c_4$.
Returning to Eq.~(\ref{eq_w3_formal}) the new model for $S_w$ finally reads 
\begin{eqnarray}  \label{eq_Sw_new}
 S_w \! &\!\! = \!\! &\! a_6\, \left(1 - \frac{2\,c_6\ell  K^{-1/2} \tilde{N}}{1 + 2\,c_6\ell  K^{-1/2} \tilde{N}} \right)  
                      + d_6  \ell\, \overline{w^2}^{\,-1} \frac{\partial{\overline{w^2}}}{\partial z} \,\,\, \mathrm{if}\,\, \tilde{N}  > 0,  \nonumber \\
 S_w \! &\!\! = \!\! &\! a_6 + d_6  \ell\, \overline{w^2}^{\,-1} \frac{\partial{\overline{w^2}}}{\partial z} \quad \mathrm{otherwise},  
\end{eqnarray}
with $\overline{w^2} = (2/3)\,K$, $a_6 \approx -1.5$, $0.1\lesssim c_6 \lesssim 0.2$, and $d_6 < 0$ subject to further
analysis with 3D RHD simulations. The closure model for $\overline{w^3}$ is hence suggested to read
\begin{eqnarray}  \label{eq_w3_new}
 \overline{w^3} \! &\!\! = \!\! &\! a_6\, \left(1 - \frac{2\,c_6 \ell K^{-1/2} \tilde{N}}{1 + 2\,c_6 \ell K^{-1/2} \tilde{N}}\right)  \overline{w^2}^{3/2}\!
                      + d_6  \ell\, \overline{w^2}^{\,1/2} \frac{\partial{\overline{w^2}}}{\partial z} \,\,\, \mathrm{if}\,\, \tilde{N}  > 0, \nonumber \\
 \overline{w^3}  \! &\!\! = \!\! &\! a_6 \overline{w^2}^{3/2}\!
                     + d_6  \ell\, \overline{w^2}^{\,1/2} \frac{\partial{\overline{w^2}}}{\partial z}  \quad \mathrm{otherwise},  
\end{eqnarray}
with $a_6$, $c_6$, and $d_6$ taken as for $S_w$. If the time scale $\tau=2\,K/\epsilon$ is available instead of $\ell$, as in 
comparisons with numerical simulations, one may use the relations from \citet{Kupka2022},
\begin{equation}    \label{eq_rewrite_K_tau}
  c_4\ell  K^{-1/2} \tilde{N} = (c_3/(2\,c_2))\,(\tau/\tau_{\rm b}) = 0.078125\,(\tau/\tau_{\rm b})
\end{equation}
and
\begin{equation}   \label{eq_rewrite_K_w2}
  c_4\ell  K^{-1/2} \tilde{N} =  c_4\,\sqrt{2/3}\,\ell\,\overline{w^2}^{\,-1/2} \tilde{N} =  c_4^{\prime}\,\ell\,\overline{w^2}^{\,-1/2} \tilde{N}
\end{equation}
with $c_4\,\sqrt{2/3}  =  c_4^{\prime} \approx  0.16$, to rewrite the equations and compute $S_w$ or $\overline{w^3}$ from
Eq.~(\ref{eq_Sw_new}) or Eq.~(\ref{eq_w3_new}). With Eq.~(\ref{eq_w2t_Z99}) for $\overline{w^2\theta}$ and 
Eq.~(\ref{eq_wt2_with_grad}) for $\overline{w\theta^2}$ and Eq.~(\ref{eq_w3_new}) for $\overline{w^3}$ a closed set 
of equations for these three TOMs in terms of second order moments and mean structure quantities was obtained. 
This new TOM model can be used to close three-equation non-local convection models such as that one of \citet{kuhfuss1987} 
(see also \citealt{Kupka2022} and \citealt{Ahlborn2022}). Introducing $D_4 = -d_4\,\ell\,\overline{w^2}$
and $D_5 = -d_5\,\ell\,\overline{w^2}$ one  can rewrite Eq.~(\ref{eq_w2t_Z99}) and Eq.~(\ref{eq_wt2_with_grad}) into
\begin{equation}   
\overline{w^2\theta} = a_1\, \overline{w^3}\, \left(\overline{w^2}\right)^{-1}\, \overline{w\theta}  
                                   +d_4\, \ell\,\overline{w^2}\, \frac{\partial{\overline{w\theta}}}{\partial  z},  \label{eq_w2t_homogenized}
\end{equation}
\begin{equation}
    \overline{w\theta^2} = a_5\,  \overline{w^3}\, \left(\overline{w^2}\right)^{-2}\, \left(\overline{w\theta}\right)^2
                                        +d_5\, \ell\,\overline{w^2}\, \frac{\partial{\overline{\theta^2}}}{\partial  z},  \label{eq_wt2_homogenized} 
\end{equation}
where $a_1 = 1$, $a_5$ is in the range of $1$ to $2$, and $d_4 < 0$ as well as $d_5 < 0$ like $a_5$ have to be 
determined from comparisons with 3D numerical simulations of convection. Appendix~\ref{app_KF_model} shows how
to use these results to close the convection model of \citet{kuhfuss1987}.

\section{Testing the model}   \label{sect_tests}

The new TOM model consists of Eq.~(\ref{eq_w3_new}), (\ref{eq_w2t_homogenized}), and~(\ref{eq_wt2_homogenized})
for $ \overline{w^3}$, $\overline{w^2\theta}$, and $\overline{w\theta^2}$ and relations for
the skewness of vertical velocity $S_w$ and temperature $S_{\theta}$, Eq.~(\ref{eq_Sw_new}) and~(\ref{St_closure}).
This new model was tested by means of 3D RHD simulation data and compared to several closure models from the literature.
The model was required to agree with direct computations of the third order moments from the 3D RHD data when its closure 
expressions were evaluated with input data from the same source. A perfect closure would yield results  nearly identical to a direct 
 computation of the three TOMs as well as $S_w$ and $S_{\theta}$. This way the predictive capabilities of the new closure 
model were probed independently from other approximations made within standalone non-local 
convection models such as that one of \citet{kuhfuss1987}. The comparison models included,
firstly, the DGA given by Eq.~(\ref{eq_w3_DGA}), (\ref{eq_w2t_DGA}), and~(\ref{eq_wt2_DGA}), and secondly, the model 
Eq.~(\ref{eq_wt2_GH2002}) for $\overline{w\theta^2}$ from \citet{gryanik02b} combined with the model of
\citet{zilitinkevich99b}, Eq.~(\ref{eq_w2t_Z99}), for $\overline{w^2\theta}$, which for convenience is referred to
here as the TSMF model with DGA. The TSMF model was also tested without DGA terms ($D_4=0$ and 
$D_5^{\prime}=0$), as suggested by several authors before (see Sect.~\ref{sect_existing}). Eq.~(\ref{eq_w2t_Z99}) 
and~(\ref{eq_wt2_GH2002}) cannot be used self-consistently in non-local convection models unless 
$S_w$ and $S_{\theta}$ are known from some other source. To benchmark the impact of Eq.~(\ref{eq_w3_new})
on the predictions of Eq.~(\ref{eq_w2t_homogenized}) and~(\ref{eq_wt2_homogenized}), in the TSMF model with 
and without DGA terms the quantities $S_w$ and $S_{\theta}$ were taken directly from the 3D RHD data. 

\begin{figure*}[t]
\centering
\includegraphics[width=0.45\textwidth]{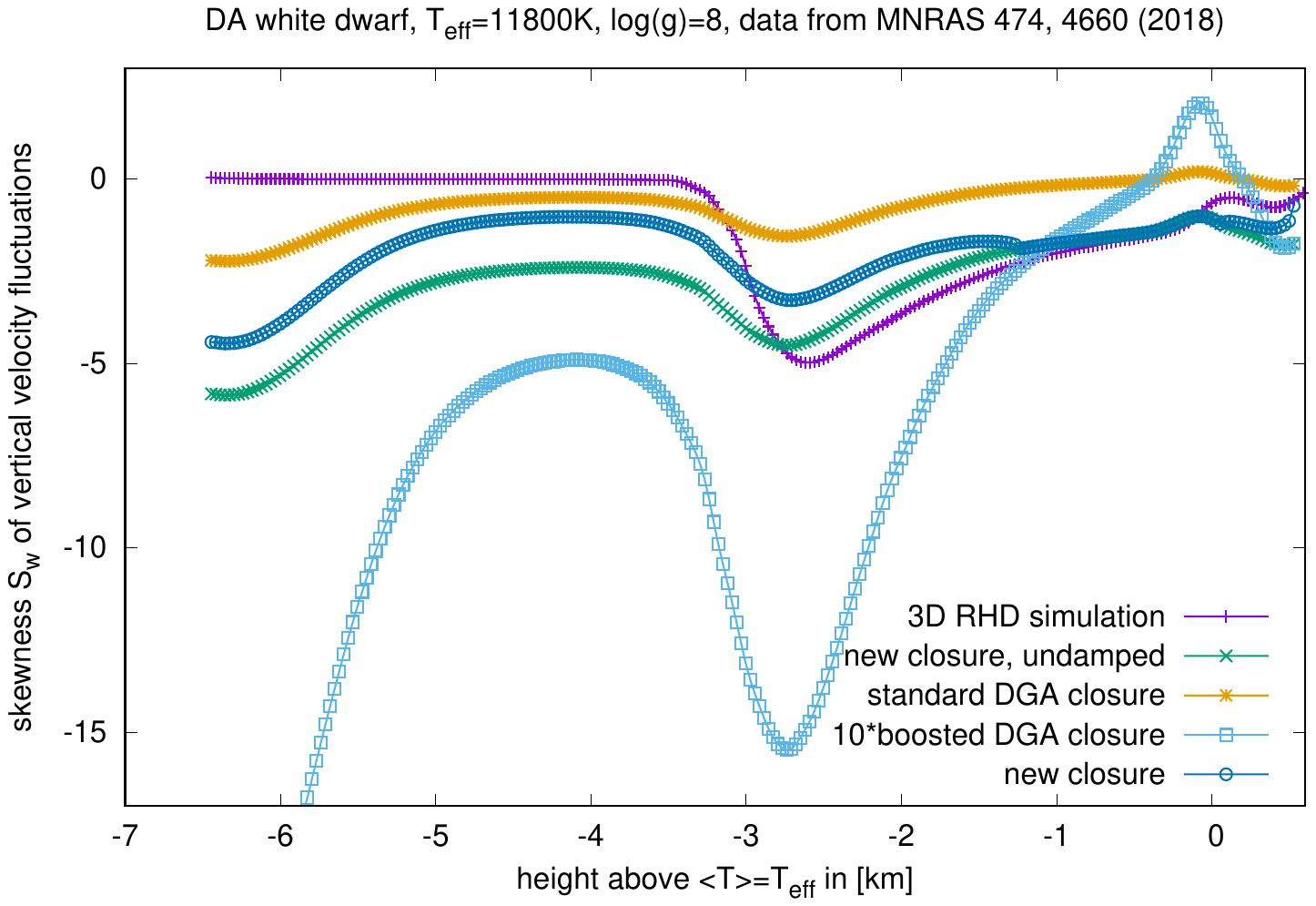}
\includegraphics[width=0.45\textwidth]{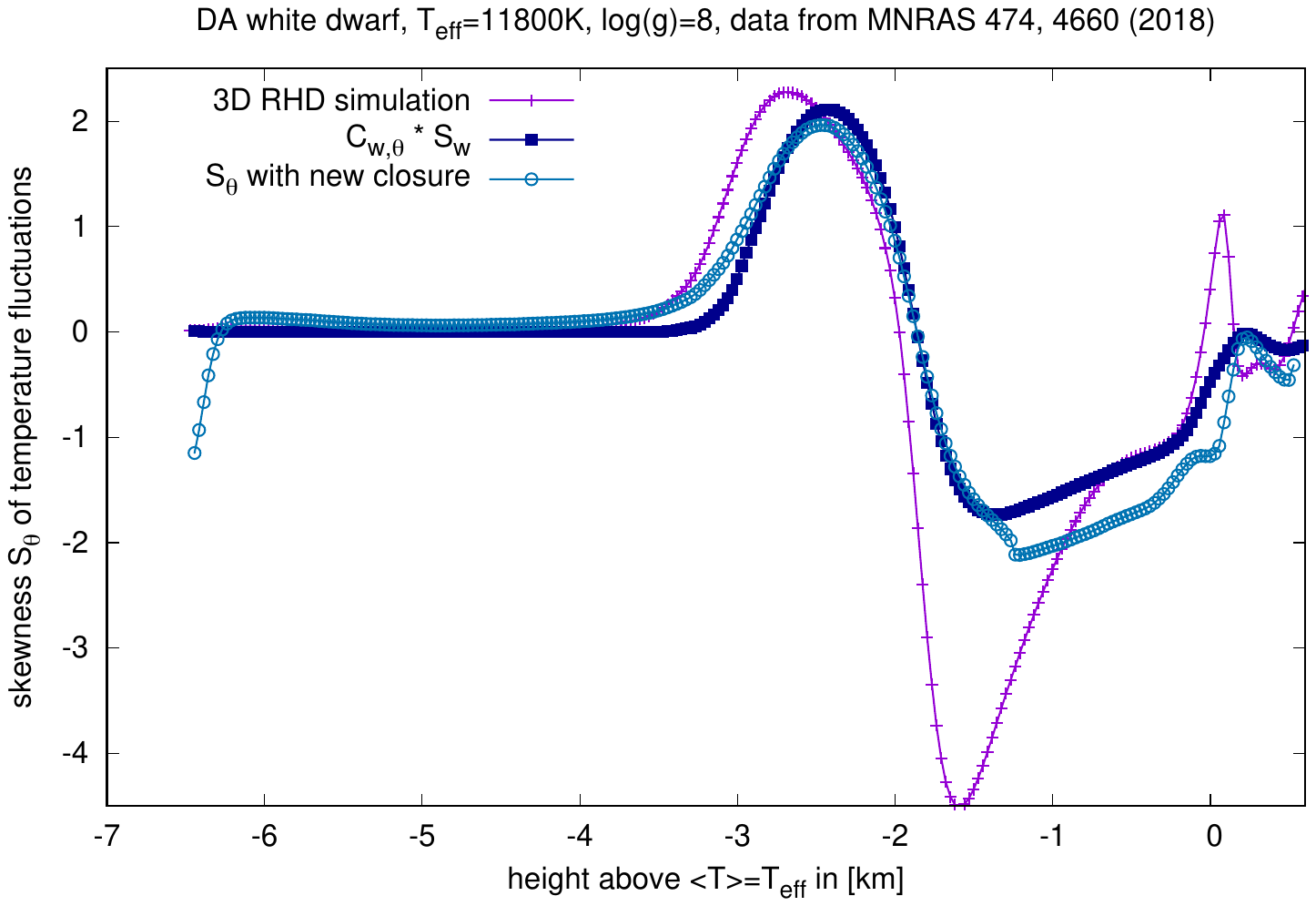}
\includegraphics[width=0.45\textwidth]{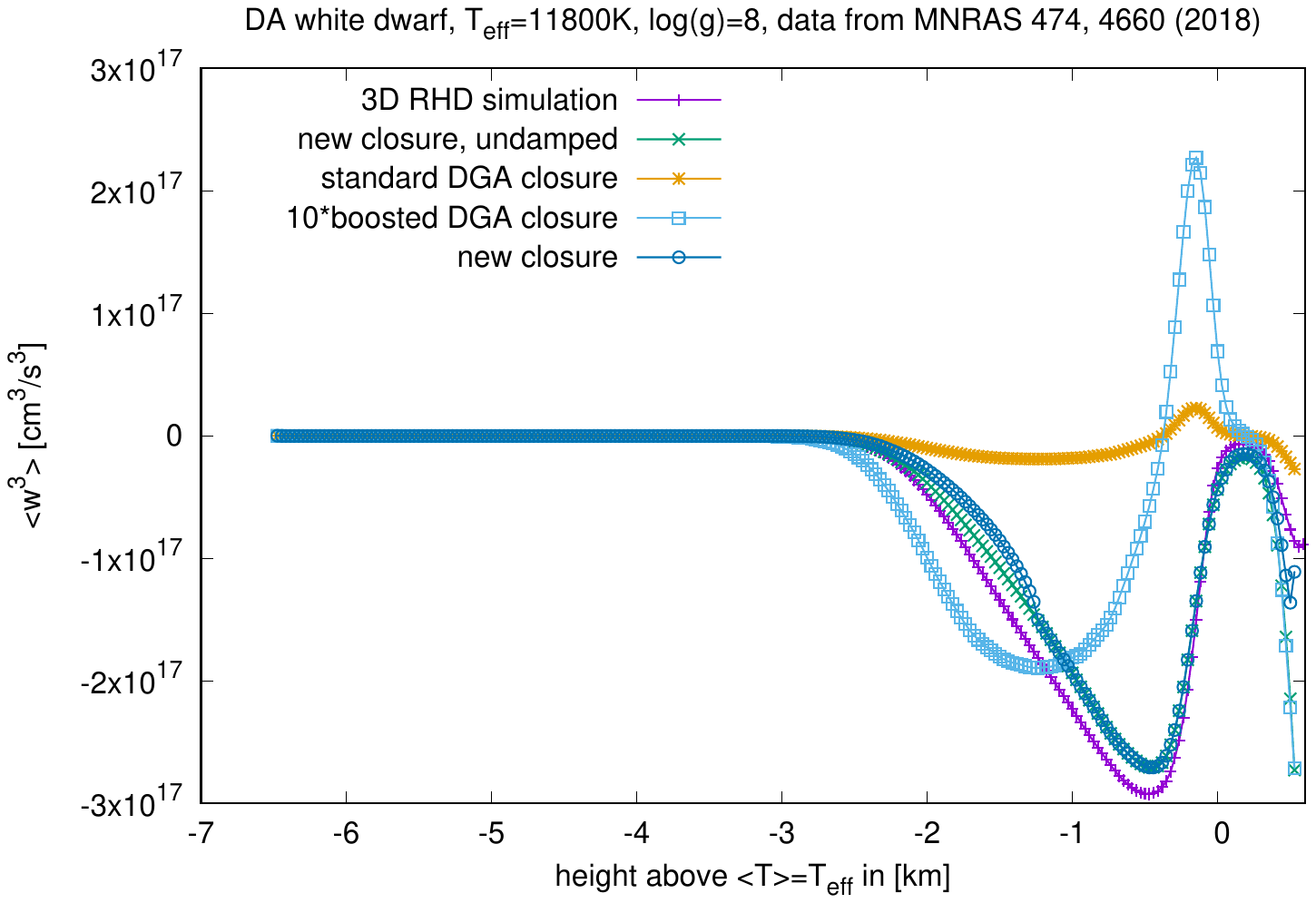}
\includegraphics[width=0.45\textwidth]{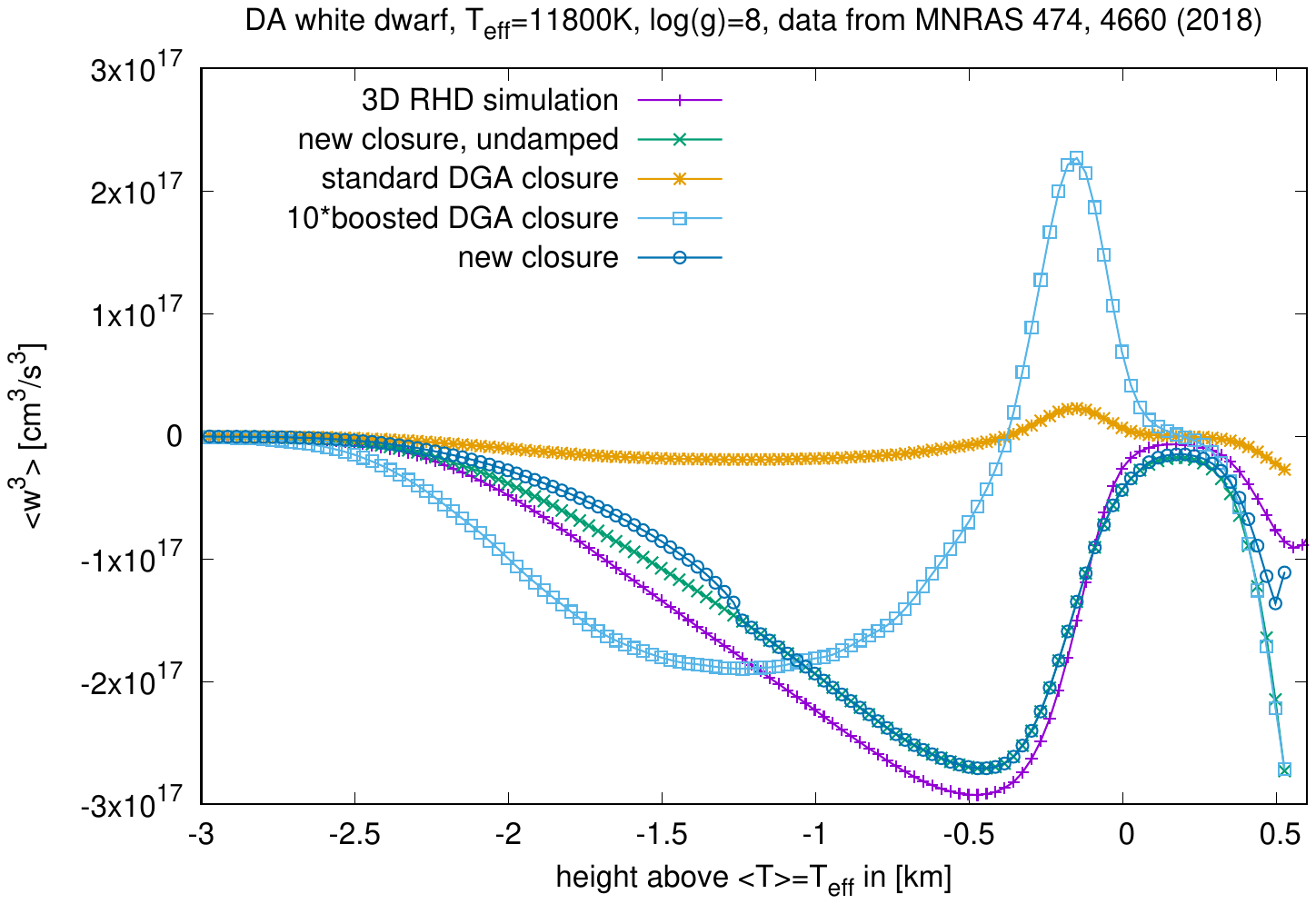}
\caption{The skewnesses $S_w$ and $S_{\theta}$ and the TOM  $\overline{w^3}$ in
              a 3D RHD simulation of convection in the upper envelope of a DA white dwarf.
              The upper row of panels compares $S_w$ (left panel) and $S_{\theta}$ (right panel)
              for the closure models discussed in Fig.~\ref{Fig1} with input data taken from the
              3D RHD simulation. The colour coding is the same as for Fig.~\ref{Fig1} 
              and~\ref{Fig2}. The lower row of panels displays a test of
              these closure models for the TOM  $\overline{w^3}$ (left panel) alongside a zoom into the upper half
              of the simulation domain (right panel). A detailed discussion is given in the main text.
\label{Fig4}
}
\end{figure*}

For the test three already published numerical simulations were used: the first data set was taken 
from a simulation of solar surface convection with closed upper and lower vertical boundaries 
\citep{kupka07b} performed with the code of \citet{robinson03b}. This simulation included only a rather 
shallow solar photosphere. Its closed vertical boundaries directly affect predictions of the behaviour 
of $S_w$, $S_{\theta}$, and the TOMs as a function of depth in the simulation domain. When compared
to a simulation with vertically open boundary conditions this allowed quantifying their role.
The model assumed a $T_{\rm eff}$ of 5777~K and a $\log g$ of 4.4377. The chemical composition 
is  described in \citet{kupka07b}. The predictions of $S_w$ of the different TOM models for this simulation 
are shown in Fig.~\ref{Fig1} and further results are provided in Appendix~\ref{app_comp_FJR}. The second 
data set was obtained from a simulation of solar surface convection performed with the code of \citet{Muthsam10a} 
presented in \citet{belkacem19b}. It featured open vertical boundary conditions and a photosphere 
ranging almost 600~km above the layer where the average temperature equals the solar $T_{\rm eff}$ 
of 5777~K. It also assumed a $\log g$ of 4.4377. Details on the chemical composition 
and the $T_{\rm eff}$ measured from the simulation, found to be more close to 5750~K, were given 
in \citet{belkacem19b}. Due to the limited accuracy of the TOM models such differences were not important 
for the test described below. Finally, 3D RHD simulation data of a DA white dwarf with a $T_{\rm eff}$ close 
to 11800~K and a $\log g$ of 8 with a pure hydrogen composition were taken from a data set discussed in
\citet{kupka18b}. In that simulation, performed with the code of \citet{Muthsam10a}, the entire convective 
zone including its neighbouring overshooting regions was contained completely inside the simulation domain.

\begin{figure*}[t]
\centering
\includegraphics[width=0.45\textwidth]{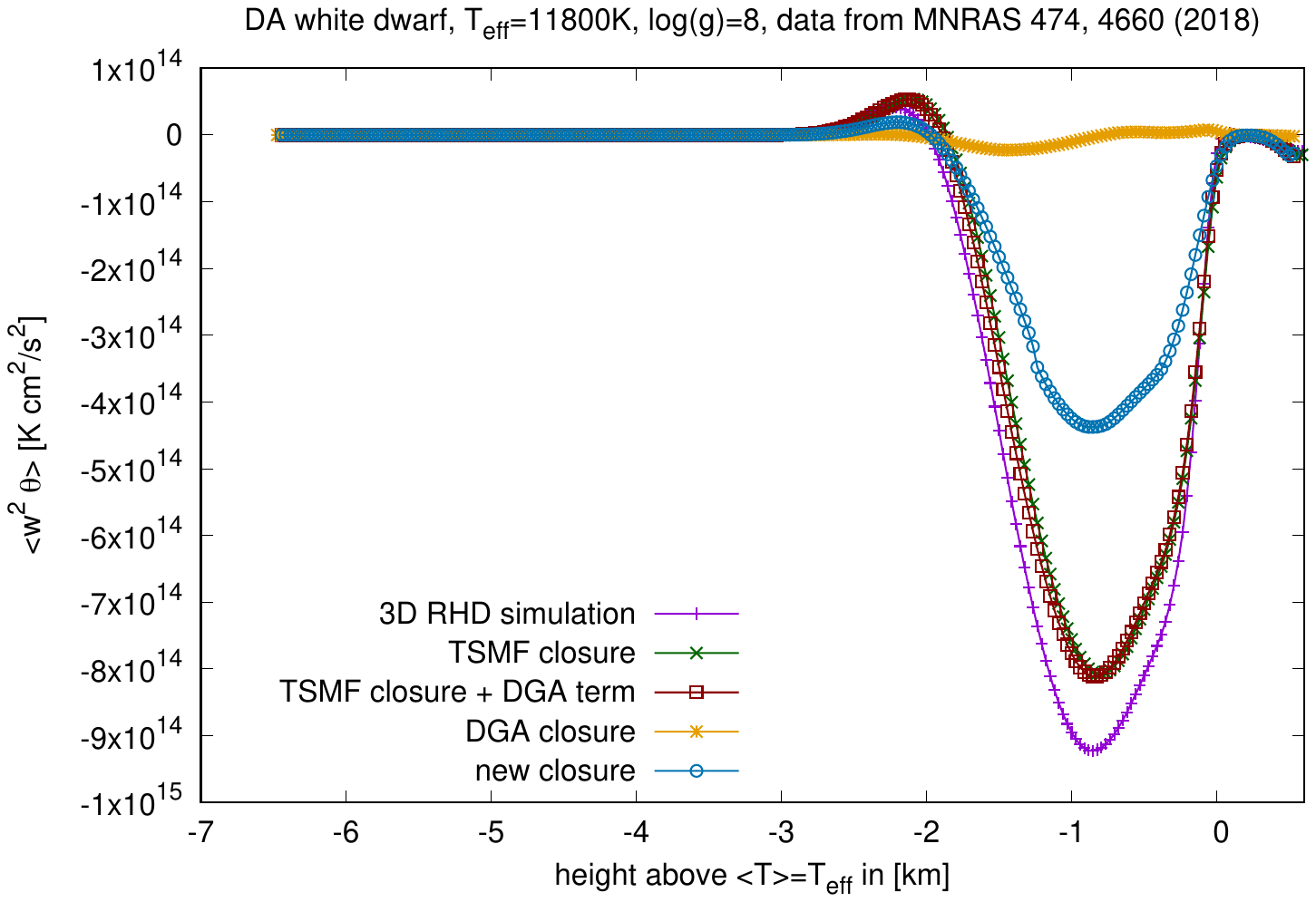}
\includegraphics[width=0.45\textwidth]{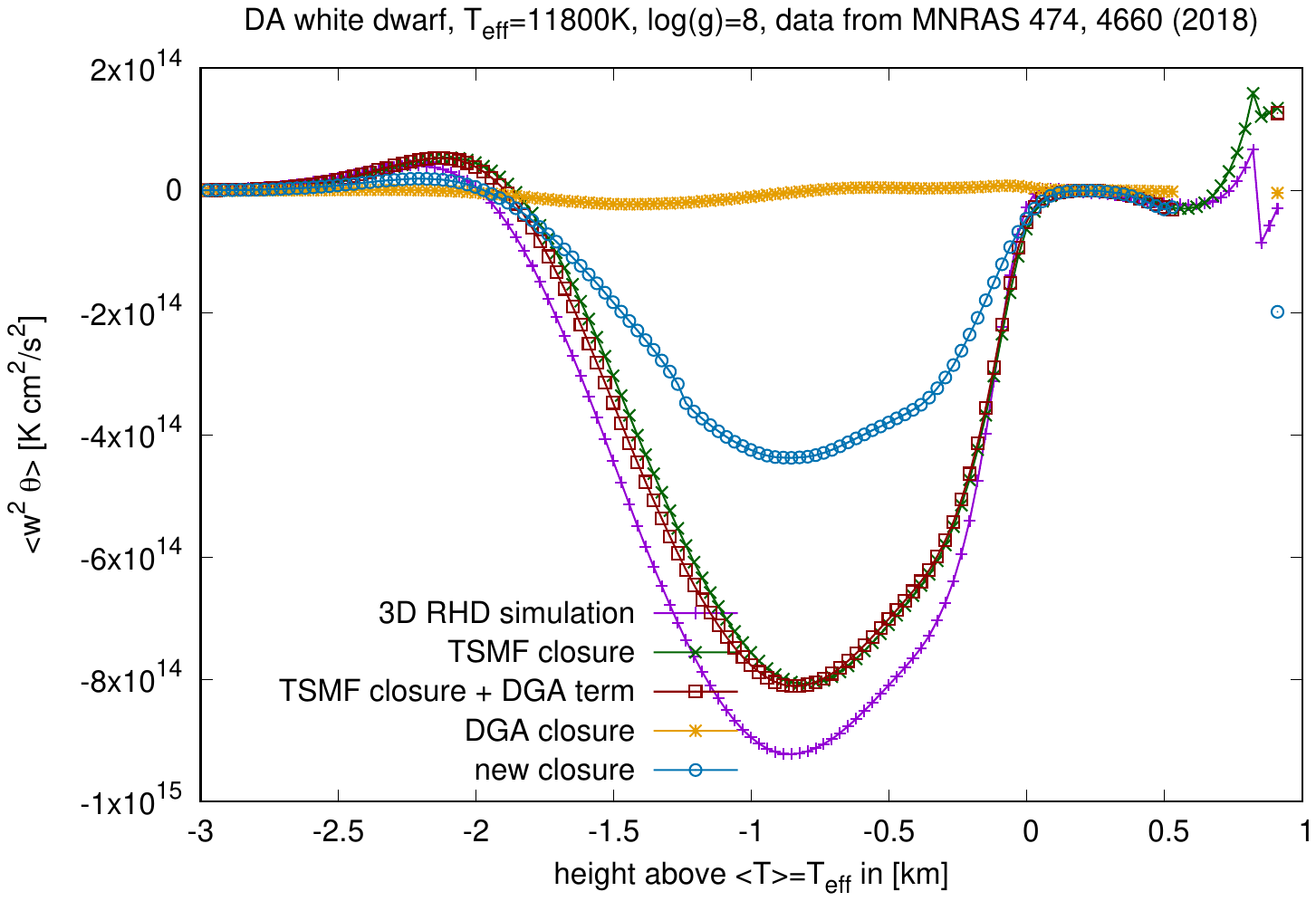}
\includegraphics[width=0.45\textwidth]{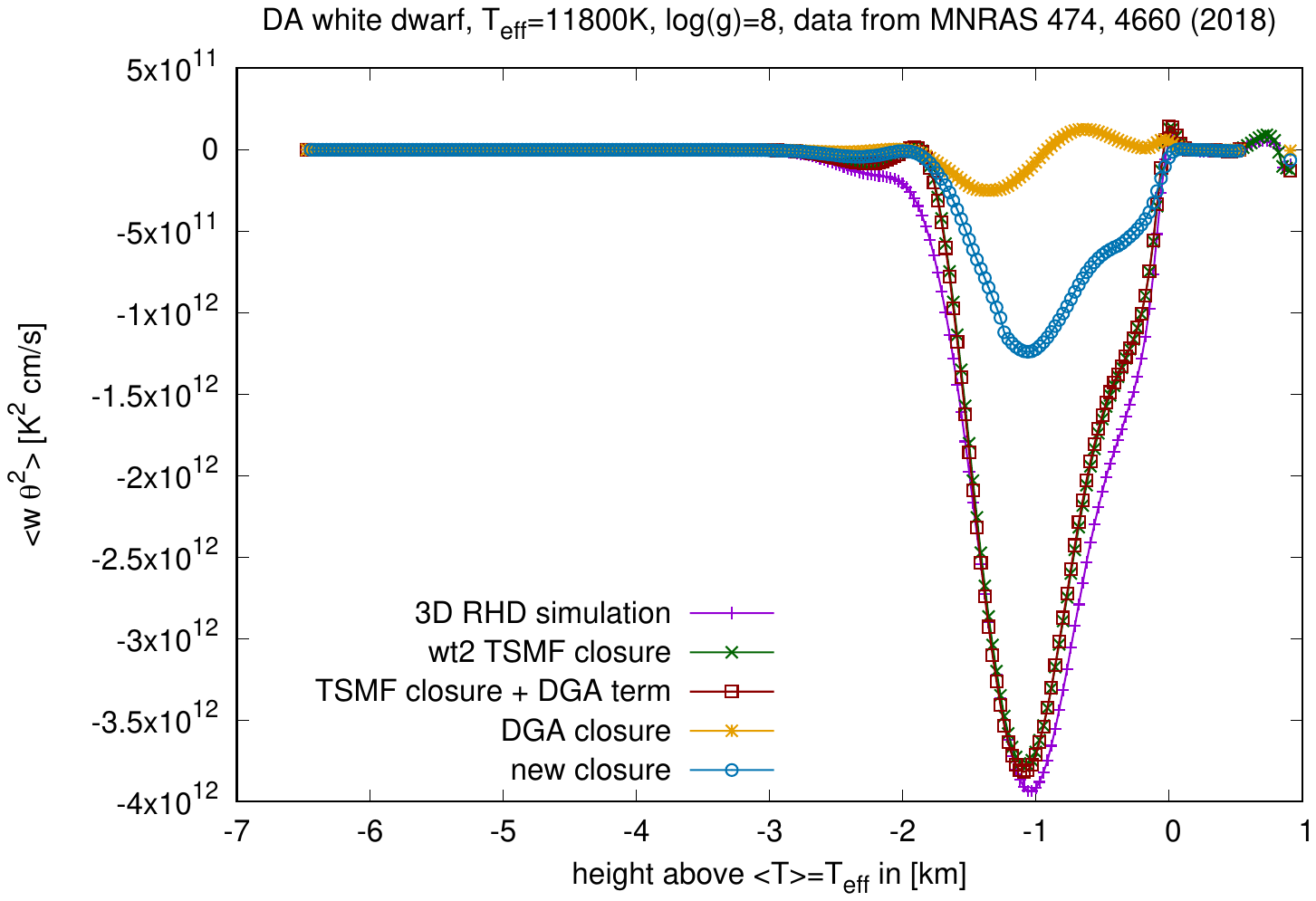}
\includegraphics[width=0.45\textwidth]{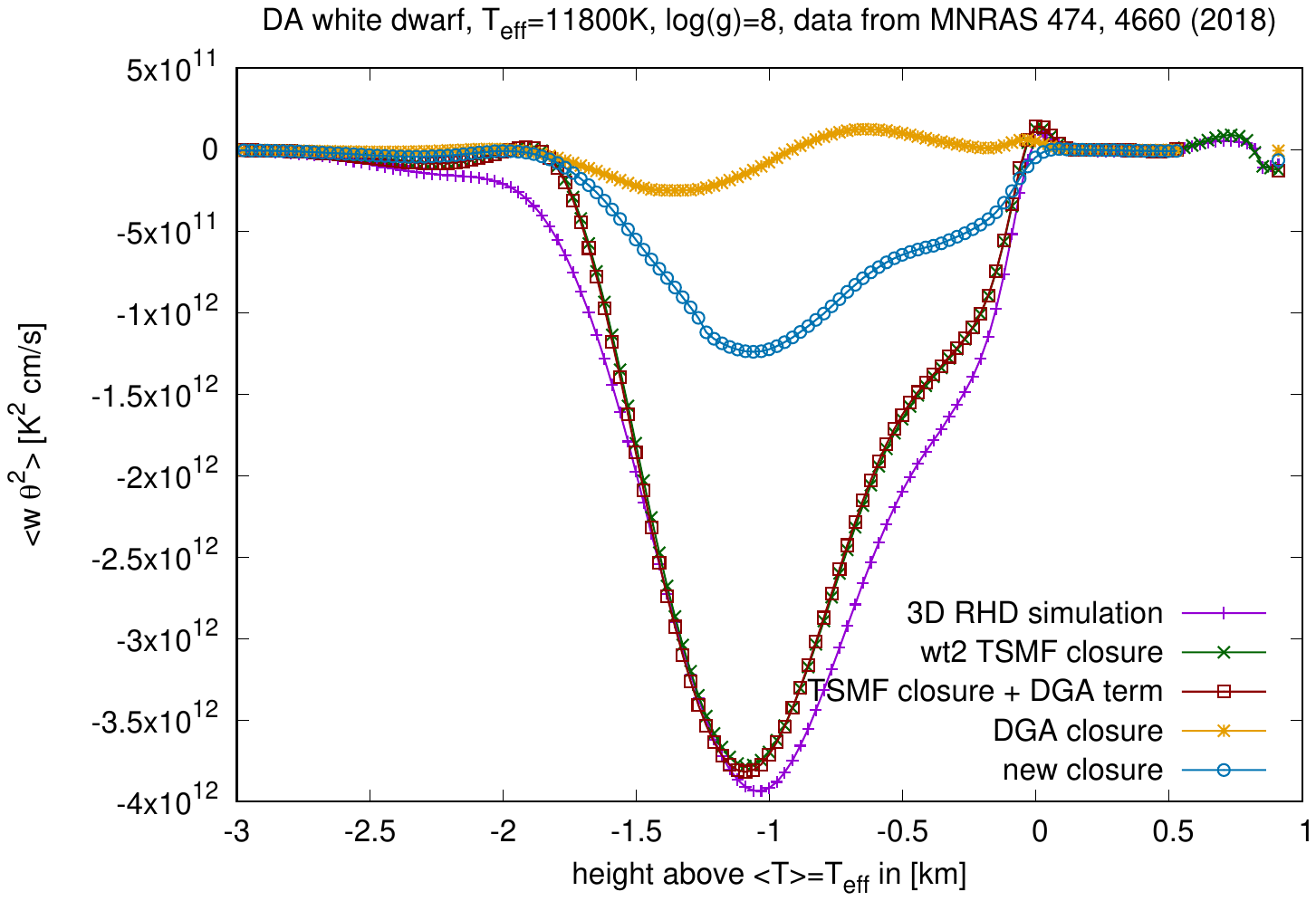}
\caption{Testing closure models for the TOMs $\overline{w^2\theta}$ and $\overline{w\theta^2}$ with
              data of a 3D RHD simulation of envelope convection in a DA white dwarf. The same models
              as discussed in Figs.~\ref{Fig1} and~\ref{Fig3} were investigated. The case of the TOM $\overline{w^2\theta}$ 
              is shown in the upper row for the full vertical range (left panel) and for the upper half of the domain (right panel). 
              For line style and colour coding see Fig.~\ref{Fig3}. The lower row repeats this comparison for $\overline{w\theta^2}$. 
              A discussion of the results is given in the main text.
\label{Fig5}
}
\end{figure*}

The closure parameters of the different TOM models were first roughly adjusted according to the
simulation of \citet{kupka07b} and then refined from tests with the data of \citet{belkacem19b}. The data of 
\citet{kupka18b} was not considered in this process except for $c_6$ for which the decision was made 
to perform all the comparisons with a smaller value than implied by the solar simulations to avoid overestimated 
damping  of the skewness in the extended Deardorff countergradient layer which appears in the lower overshooting 
region of the DA white dwarf. After this process the values for these parameters were not adjusted any 
further. The values used for all the results shown in Fig.~\ref{Fig1}--\ref{Fig5} are summarized 
in the following. For the DGA, the diffusivities had been defined as $D_1 = -d_1\,\ell\,\overline{w^2}$, 
$D_2 = -d_2\,\ell\,\overline{w^2}$ as well as $D_3 = -d_3\,\ell\,\overline{w^2}$. Having applications to stellar 
evolution calculations in mind $\ell$ was computed similarly to \citet{Kupka2022} and \citet{Ahlborn2022}
from using $\tau  =  2\,K/\epsilon$ and  $\epsilon = c_{\epsilon}\,K^{3/2} /\ell$ with $c_{\epsilon}=0.8$ and 
$\Lambda$ taken from their model. This was not crucial, as for the cases shown in Fig.~\ref{Fig1}--\ref{Fig5} 
the alternative choice of $\Lambda = 1.5\,H_{\rm p}$ was found to yield qualitatively and quantitatively 
similar results.\footnote{This is explained by the radius 
        dependence of the model of \citet{Kupka2022} for $\Lambda$. The model predicts a strong damping 
        of $\Lambda$ in the overshooting region of convective cores, but this effect
        is much weaker near the surface of main sequence stars with a radius $R$, where $H_{\rm p} \ll R$.}
The values $d_1=-0.01$, $d_2=-0.01$, and $d_3 = -0.1$ allow positive diffusivities $D_1$, $D_2$, and $D_3$ 
to be multiplied with a negative gradient in Eq.~(\ref{eq_wt2_DGA})-(\ref{eq_w3_DGA}), as required by the 
model.\footnote{By introducing $D_i > 0$ and $d_i < 0$ the DGA model could be written in its standard 
                         notation. This simplified the comparison with the new TOM model.}
                         
As already mentioned, the TSMF model was used as a reference for the computation of $\overline{w^2\theta}$ 
and $\overline{w\theta^2}$ with the new TOM model. It was used in its original form, Eq.~(\ref{eq_w2t_CD98}) 
and Eq.~(\ref{eq_wt2_M99}), or, equivalently, Eq.~(\ref{eq_w2t_Z99}) and Eq.~(\ref{eq_wt2_GH2002}) with 
$a_1=1$, $a_2=1$, $D_4=0$, and $D_5^{\prime}=0$. It was also used with additional gradient terms (denoted 
as TSMF closure + DGA term in the following), based on Eq.~(\ref{eq_w2t_Z99}) 
and Eq.~(\ref{eq_wt2_GH2002}), where $a_1=1$, $a_2=1$ while $D_4= -d_4^{\prime}\,\ell\,\overline{w^2}$ and
$D_5^{\prime} = -d_5^{\prime}\,\ell\,\overline{w^2}$ with $d_4^{\prime}=-0.04$  and $d_5^{\prime}=-0.01$. 
Especially the choice of $d_4^{\prime}$ improved model predictions at the solar surface.

For the new TOM model a value of $a_4=1.5$ was used for the computation of $S_{\theta}$ 
according to Eq.~(\ref{St_closure}). This is right in the middle of the suggested range discussed 
in Sect.~\ref{sect_Stheta}. Taking $a_2=1$ the relation $a_5 = a_2\, a_4$ discussed in 
Sect.~\ref{sect_Stheta} would imply $a_5= 1.5$. Since Eq.~(\ref{eq_wt2_homogenized}) for the computation 
of $\overline{w\theta^2}$ in the new model also has a downgradient contribution, however, a smaller 
value of $a_5= 0.5$ turned out to yield better results while, in comparison with Eq.~(\ref{eq_wt2_DGA}),
the downgradient contribution was increased by setting $d_5 = -0.06$. This accounts 
for changes introduced by the evaluation of Eq.~(\ref{eq_wt2_homogenized}) when $\overline{w^3}$ 
is computed from Eq.~(\ref{eq_w3_new}) instead of taking it directly from the 3D RHD data, which on
the contrary has to be done for the TSMF model. The coefficients in Eq.~(\ref{eq_w2t_homogenized}) 
for the computation of $\overline{w^2\theta}$ with $\overline{w^3}$ taken from Eq.~(\ref{eq_w3_new}) 
were set to $a_1=0.6$  and $d_4=-0.06$, a similar change as in the case of $\overline{w\theta^2}$.
The solutions based on the new TOM model were all computed by means of Eq.~(\ref{eq_w3_new}) with 
$a_6=-\sqrt{2}$  and $d_6=-0.2$. The choice of $a_6$ was motivated by the realizability constraint 
for the kurtosis of vertical velocity, $K_w \geqslant 1 + S_w^2$, for skewed flow with a normal distribution,
$K_w=3$ (see \citealt{kupka07b}). The extra damping by the harmonic limiter required $c_6  = 0.1$. 
Values closer to 0.2 improved the match for the 3D RHD in the solar photosphere but increased the differences 
in the lower countergradient layer of the DA white dwarf. Eq.~(\ref{eq_rewrite_K_tau}) was used to 
simplify comparisons with different computations for $\tau$ in 3D RHD numerical simulations. If the new 
TOM model were used in a non-local convection model with an isotropic velocity field, model coefficients might 
have to be scaled according to Eq.~(\ref{eq_rewrite_K_w2}).

For Fig.~\ref{Fig1} $S_w$ was computed with the new closure with ($c_6 = 0.1$) as well as without ($c_6 = 0$) 
damping and compared to the DGA, a DGA solution with ten times larger diffusivity, and a direct computation
from the 3D RHD simulation, each for the solar granulation simulations presented in \citet{belkacem19b} and
in \citet{kupka07b} (colour and line coding are described in the figure caption and were used also for the
other figures). Clearly, the DGA completely failed to predict $S_w$ and even if $d_3$ were boosted from
$-0.1$ to $-1$, the better match inside the quasi-adiabatic convective zone were traded with a failure
in size and sign at the top of the convective zone and in the photosphere. The closed boundary at the
bottom of the 3D RHD by \citet{kupka07b} led to somewhat larger discrepancies, but this is uncritical,
since the solid plate boundary condition was not considered in the design of the models.

Fig.~\ref{Fig2} compares $S_{\theta}$ from the new model and predictions for $\overline{w^3}$ from the same 
set of models as in Fig.~\ref{Fig1} with 3D RHD simulation data from \citet{belkacem19b}. For $S_{\theta}$,  
Eq.~(\ref{St_closure}) with $a_4=1$ and $C_{w\theta}\, S_w$ taken from the 3D RHD data was compared with
a direct evaluation of the 3D RHD simulation and with a calculation based on the new closure Eq.~(\ref{eq_Sw_new}) 
to compute $S_w$ assuming $a_4=1.5$ for evaluating Eq.~(\ref{St_closure}). The latter provided an at least qualitatively 
acceptable approximation which, however, missed the maximum at the top of the convective zone. Differences 
at the bottom could also have been caused by properties of the boundary conditions. The prediction of $\overline{w^3}$ 
by the new TOM model was satisfactory also from a quantitative viewpoint contrary to the DGA which missed 
the numerical range of values in the interior of the convective zone by an order of magnitude and at the top of 
the convective zone even mispredicted the sign of $\overline{w^3}$, a problem only worsened when boosting $d_3$.

In Fig.~\ref{Fig3} the predictions of the cross-correlations $\overline{w^2\theta}$ and $\overline{w\theta^2}$ are 
compared for the same models as for the case of $\overline{w^3}$ in Fig.~\ref{Fig2}, except that no 
boosted DGA solutions are presented, because they drastically fail anyway. Instead, the TSMF model with and 
without DGA term are shown. The data was again taken from the simulation of \citet{belkacem19b}. The DGA 
achieved acceptable results only  for the top of the convective zone and the adjacent overshooting region. Inside the 
quasi-adiabatic interior of the convection zone, however, it failed to predict $\overline{w^2\theta}$ by an order of
magnitude and $\overline{w\theta^2}$ by a factor of three. The new model provided a much better prediction 
of both $\overline{w^2\theta}$ and $\overline{w\theta^2}$ which recovered the sign and the shape of these 
quantities as a function of depth. For the TSMF model with $S_w$ and $S_{\theta}$ taken from the 3D RHD 
simulation the variant with a DGA contribution allowed a slightly better prediction of $\overline{w^2\theta}$
than the new model. Without this contribution the TSMF failed to predict the correct sign of this quantity in the 
solar photosphere and  performed worse than the new model. The DGA contribution to the TSMF model turned
out to be much less important for $\overline{w\theta^2}$. 

The white dwarf simulation of \citet{kupka18b} provided an alternative test scenario which also included 
an extended countergradient layer and an overshooting zone underneath the convective zone. 
Fig.~\ref{Fig4} compares $S_w$, $S_{\theta}$, $\overline{w^3}$, and a zoom into the upper half of the 
simulation domain for a more detailed analysis of the modelling of $\overline{w^3}$. The DGA, independently
of whether $d_3=-0.1$ or $d_3 = -1$, resulted in very poor predictions of $S_w$ and $\overline{w^3}$. 
On the contrary, the new model recovered $\overline{w^3}$ very well. In the upper photosphere damping through
setting $c_6=0.1$ improved the comparison of the new TOM model with a direct computation of $\overline{w^3}$ 
over the choice $c_6=0$ whereas in the countergradient layer (between $-1.3$~km and $-2$~km for the chosen 
zero point of vertical depth) a weaker or no damping appears to be preferred. The results for $S_w$ indicate
that the lengthscale $\Lambda$ used in the computation of the diffusivities of the downgradient contributions 
to $\overline{w^3}$ should drop more rapidly to allow $S_w \rightarrow 0$ in the wave dominated region
(see also \citealt{kupka18b}). Further improvements were also found to be necessary to improve predictions
of $S_{\theta}$ in the countergradient region. The results on the cross-correlations $\overline{w^2\theta}$ and 
$\overline{w\theta^2}$ shown in Fig.~\ref{Fig5} confirmed the conclusions drawn from the solar case (Fig.~\ref{Fig3}).
The DGA was not able to successfully predict their functional shape, magnitude, and sign. The new model
achieved this goal to a much better extent. Even a match comparable with the TSMF model and the 3D RHD 
simulation would have been possible by choosing $a_1=1.2$ instead of $0.6$ and  $a_5=1.5$ instead of $0.5$ 
due to a much more realistic depth dependence of the new TOM model.

\section{Discussion}   \label{sect_discussion}

Challenging its widespread use to compute the TOMs  $\overline{w^3}$, $\overline{w^2\theta}$, and $\overline{w\theta^2}$,
the DGA was probed with 3D RHD simulations of solar surface convection and a 3D RHD simulation of convection in the upper 
envelope of a DA white dwarf. Simulations of this kind had already passed a large variety of tests (see \citealt{nordlund09r}, 
\citealt{kupka17b}, \citealt{kupka18b}, \citealt{belkacem19b}) which made them suitable for challenging closure models. The DGA 
was found to perform rather poorly in both cases, as it agreed with the 3D RHD simulations only near the boundary of convective 
regions and in neighbouring overshooting layers. In the interior of the convection zones the DGA was found to differ by factors 
of three  to ten from the simulation data and it even mispredicted the sign of $\overline{w^3}$ at the stellar surface including
a positive flux of kinetic energy,  $F_{\rm kin} > 0$, while from 3D RHD simulations this quantity was found to remain negative in 
the stellar photosphere, $F_{\rm kin} < 0$, unless forced to do so by a closed upper boundary condition (see Appendix~\ref{app_fkin}).

The new TOM model developed in Sect.~\ref{sect_model} was found to allow a quantitative improvement over the DGA by 
up to an order of magnitude. The model also qualitatively improved the predictions of the TOMs $\overline{w^3}$, 
$\overline{w^2\theta}$, and $\overline{w\theta^2}$ by recovering, in most cases, their correct dependency as functions 
of depth and their correct sign. The same held for the skewness $S_w$,  even more so, when considering the large mismatch
of the DGA for these quantities. Encouragingly, the same qualitative and quantitative improvements were found for both of the two
rather different scenarios, solar surface granulation and shallow envelope convection in a DA white dwarf (Figs.~\ref{Fig1}--\ref{Fig5}).
Further improvements might be required for the modelling of $S_{\theta}$, where an additional contribution in the countergradient
layer (Fig.~\ref{Fig4}, upper right panel) appears to be missing, and to achieve a more efficient damping of the scalelength 
$\Lambda$ at the bottom of the lower overshooting zone (Fig.~\ref{Fig4}, upper left panel). The latter would ensure 
that $S_w \rightarrow 0$ in the wave-dominated region. The modelling of $\overline{w\theta^2}$ could also benefit
from such potential improvements, since for the DA white dwarf simulation the latter was underestimated in magnitude 
where $|S_{\theta}|$ was underestimated, too (see Figs.~\ref{Fig4} and \ref{Fig5}).

The TSMF model was found to provide an even better approximation to the 3D RHD data, at least  when extended with 
a DGA contribution as suggested by \citet{zilitinkevich99b} and \citet{gryanik02b}. This was expected, since the new TOM 
model is based on approximations to $S_w$ and $S_{\theta}$ rather than taking their values directly from simulation data. 
For that same reason the TSMF model, however, cannot be used on its own in second order closure non-local models of 
turbulent convection such as those by \citet{kuhfuss1987} and \citet{Kupka2022}. Models for $\overline{w^3}$ 
to predict $S_w$ for the TSMF model which are based on the DGA or on more complete models that can be expressed 
as linear combinations of gradients of second order moments \citep{canuto98b} or, additionally, third order moments 
\citep{gryanik02b}, underestimate $S_w$ because of the similar shape of all the gradients of the second order moments 
which prevents matching $\overline{w^3}$ and $S_w$ simultaneously in the deep interior and at the convective zone
boundary (see Appendix~\ref{app_comp_FJR} and~\ref{app_fkin}). This shortcoming of gradient based models 
for $S_w$ had already been observed in \citet{kupka07d} and \citet{kupka07f} for direct numerical simulations of compressible 
convection and the conclusions on it do not change if density weighted Favre averages are used instead of plain Reynolds 
averages: their small difference in computations of $\overline{w\theta^2}$ for solar surface granulation and envelope convection 
in DA white dwarfs was demonstrated in \citealt{kupka17b} (see their Figs.~8 and~9). Moreover, in solar granulation simulations 
the Favre averaged expression for $F_{\rm kin}$ (see Appendix~\ref{app_fkin}) features the same plateau as $\overline{w^3}$ 
(Fig.~\ref{Fig1}). The purely algebraic contributions contained in the new TOM model are hence a likely explanation for its
improved performance described in Sect.~\ref{sect_tests}.

It has to be emphasized here that the closure constants of the new TOM model, by lack of a more
general theory from which it can be derived, should not be taken as an incentive to ``fit'' any kind  of astrophysical 
or other data ``better'' than it is possible with the values suggested in Sect.~\ref{sect_tests}. The concept proposed here 
is to calibrate them for a single case as in Sect.~\ref{sect_tests} and consider discrepancies to observational data rather 
as an incentive to further improvements to the functional form of the model.

It should also be noted that the seemingly ``better'' performance of the DGA model
in the wave dominated region below the overshooting zone of the DA white dwarf
envelope convection zone (Fig.~\ref{Fig4}, upper left panel) is solely due to the not
sufficiently damped scale length $\Lambda$ in the gradient diffusivity appearing in both
Eq.~(\ref{eq_w3_DGA}) and Eq.~(\ref{eq_w3_new}) and the numerical values of $d_3$ and 
$d_6$ suggested in Sect.~\ref{sect_tests}. Improving the model for $\Lambda$ should
solve this problem just as the discrepancies left for $\overline{\theta^3}$ and $\overline{w\theta^2}$
are best addressed by further improving the model for $S_{\theta}$.

\section{Conclusions}   \label{sect_conclusions}

The accurate computation of non-local convective transport is one of the greatest challenges 
in building statistical models of solar and stellar structure and evolution \citep{Kippenhahn2012}. 
Such models are equally valuable to helio- and asteroseismology \citep{Aerts2010,houdek15b}.
They are also very important in oceanography and meteorology \citep{canuto09b,mironov09b}.
Available models for the key quantities describing these processes, the third order moments of (vertical)
velocity and temperature (or entropy), were found to usually have only a limited region of applicability
(see the discussions in \citealt{losch04b,cheng05b,cai18b}). For deep convective envelopes
none of the available TOM models that require only second order moments and mean quantities 
as input were found to be satisfactory \citep{chan96b,kupka07d}. Extending ideas underlying the
TSMF model \citep{gryanik02b} and the entropy rain model \citep{Brandenburg2016} a new model
for the skewnesses $S_w$ and $S_{\theta}$ and the TOMs $\overline{w^3}$, $\overline{w^2\theta}$, 
and $\overline{w\theta^2}$ was proposed which is applicable to close three equation non-local models 
of convection \citep{kuhfuss1987,Kupka2022} designed for stellar evolution calculations. The new TOM 
model was found to provide a major qualitative and quantitative improvement over the DGA model currently 
used for this purpose. 

Possible other scenarios to which the model could be applied are the dry planetary boundary 
layer of the Earth, convective zones in the envelopes of giant planets, or the atmosphere of terrestrial planets, 
which can also feature driving by either convective cooling or heating (see \citealt{zilitinkevich99b}
and references therein) though the influence of solid, close boundaries
may have to be considered more carefully and possibly the closure constants or at the very least 
the value of $a_6$ may have to be recalibrated (see also Sect.~\ref{sect_Sw}) on the basis of simulation
data or measurements for such physical systems. Convection in clouds
in turn would require to account for the effects of evaporation and condensation. For stellar astrophysics
the inclusion of effects of rotation on $S_w$ is likely of more interest (as explored by \citealt{Kaepylae2024}).

Next steps in model development include improvements of the modelling of the skewness $S_{\theta}$
and the associated third moments $\overline{\theta^3}$ and  $\overline{w\theta^2}$ as well as the
scale length $\Lambda$, to improve its predictive capabilities for $S_w$ and $S_{\theta}$ as well as 
to better account for anisotropy in convective flows, as summarized in Sect.~\ref{sect_discussion}, 
followed by coupling the new TOM model into a non-local model of convection for computing stellar 
envelopes and finally its implementation into a stellar evolution code and eventually to also account
for the influence of rotation on non-local transport.

\begin{acknowledgements}
The author is thankful for the hospitality provided by the Wolfgang Pauli Institute, Vienna, and acknowledges
support by the Faculty of Mathematics at the University of Vienna through offering him a Senior Research Fellow 
status. The Austrian Science Fund (FWF) has supported this research through grants P~33140-N, P~35485-N, and 
PAT1338425. Permission by F.J.~Robinson to use his 3D simulation data for testing the new models is greatly appreciated.
\end{acknowledgements}

%
%

\bibliographystyle{aa}
\bibliography{references} 

\providecommand{\noopsort}[1]{}\providecommand{\singleletter}[1]{#1}%
\begin{thebibliography}{82}
\expandafter\ifx\csname natexlab\endcsname\relax\def\natexlab#1{#1}\fi

\bibitem[{Abdella \& McFarlane(1997)}]{Abdella97}
Abdella, K. \& McFarlane, N. 1997, J. Atm. Sci., 54, 1850

\bibitem[{Aerts {et~al.}(2010)Aerts, Christensen-Dalsgaard, \&
  Kurtz}]{Aerts2010}
Aerts, C., Christensen-Dalsgaard, J., \& Kurtz, D. 2010, Asteroseismology,
  Astron. \& Astrophys. Library (Heidelberg: Springer-Verlag)

\bibitem[{{Ahlborn} {et~al.}(2022){Ahlborn}, {Kupka}, {Weiss}, \&
  {Flaskamp}}]{Ahlborn2022}
{Ahlborn}, F., {Kupka}, F., {Weiss}, A., \& {Flaskamp}, M. 2022, A\&A, 667, A97
  (17 pp.)

\bibitem[{Arakawa(1969)}]{arakawa69b}
Arakawa, A. 1969, in Proc. WMO/IUGG Symp. Numerical Weather Prediction, Vol.
  IV, 8, Japan. Meteorological Agency, Japan, 1

\bibitem[{Arakawa \& Schubert(1974)}]{arakawa74b}
Arakawa, A. \& Schubert, W.~H. 1974, J. Atm. Sci., 31, 674

\bibitem[{{Belkacem} {et~al.}(2019){Belkacem}, {Kupka}, {Samadi}, \&
  {Grimm-Strele}}]{belkacem19b}
{Belkacem}, K., {Kupka}, F., {Samadi}, R., \& {Grimm-Strele}, H. 2019, A\&A,
  625, A20 (15 pp.)

\bibitem[{Belkacem {et~al.}(2006)Belkacem, Samadi, Goupil, \&
  Kupka}]{belkacem06b}
Belkacem, K., Samadi, R., Goupil, M.-J., \& Kupka, F. 2006, A\&A, 460, 173

\bibitem[{{Biermann}(1932)}]{biermann32r}
{Biermann}, L. 1932, Z. Astrophys., 5, 117

\bibitem[{B\"ohm-Vitense(1958)}]{bv58b}
B\"ohm-Vitense, E. 1958, Z. Astrophys., 46, 108

\bibitem[{Brandenburg(2016)}]{Brandenburg2016}
Brandenburg, A. 2016, ApJ, 832, 6 (19 pp.)

\bibitem[{Brun \& Browning(2017)}]{brun17b}
Brun, A. \& Browning, M. 2017, Liv. Rev. Sol. Phys., 14:4, 133 pages

\bibitem[{Cai(2018)}]{cai18b}
Cai, T. 2018, ApJ, 868, 12 (29 pp.)

\bibitem[{Canuto(1992)}]{canuto92b}
Canuto, V.~M. 1992, ApJ, 392, 218

\bibitem[{Canuto(1993)}]{canuto93b}
Canuto, V.~M. 1993, ApJ, 416, 331

\bibitem[{Canuto(1997)}]{canuto97b}
Canuto, V.~M. 1997, ApJ, 482, 827

\bibitem[{Canuto(2009)}]{canuto09b}
Canuto, V.~M. 2009, in Lecture Notes in Physics, Vol. 756, Interdisciplinary
  {A}spects of {T}urbulence, ed. W.~Hillebrandt \& F.~Kupka (Berlin: Springer),
  107--160

\bibitem[{Canuto(2011)}]{canuto11b}
Canuto, V.~M. 2011, A\&A, 528, A76 (9 pp.)

\bibitem[{Canuto {et~al.}(2007)Canuto, Cheng, \& Howard}]{canuto07b}
Canuto, V.~M., Cheng, Y., \& Howard, A.~M. 2007, Ocean Mod., 16, 28

\bibitem[{Canuto \& Dubovikov(1998)}]{canuto98b}
Canuto, V.~M. \& Dubovikov, M. 1998, ApJ, 493, 834

\bibitem[{Canuto {et~al.}(1994)Canuto, Minotti, Ronchi, Ypma, \&
  Zeman}]{canuto94b}
Canuto, V.~M., Minotti, F., Ronchi, C., Ypma, R.~M., \& Zeman, O. 1994, J. Atm.
  Sci., 51, 1605

\bibitem[{Chan \& Sofia(1996)}]{chan96b}
Chan, K.~L. \& Sofia, S. 1996, ApJ, 466, 372

\bibitem[{Cheng {et~al.}(2005)Cheng, Canuto, \& Howard}]{cheng05b}
Cheng, Y., Canuto, V.~M., \& Howard, A.~M. 2005, J. Atm. Sci., 62, 2189

\bibitem[{Christensen-Dalsgaard(2002)}]{JCD02}
Christensen-Dalsgaard, J. 2002, Rev. Mod. Phys., 74, 1073

\bibitem[{{Christensen-Dalsgaard}(2021)}]{JCD21}
{Christensen-Dalsgaard}, J. 2021, Liv. Rev. Sol. Phys., 18, 2

\bibitem[{{Christensen-Dalsgaard} {et~al.}(2011){Christensen-Dalsgaard},
  {Monteiro}, {Rempel}, \& {Thompson}}]{christensen2011}
{Christensen-Dalsgaard}, J., {Monteiro}, M.~J.~P.~F.~G., {Rempel}, M., \&
  {Thompson}, M.~J. 2011, MNRAS, 414, 1158

\bibitem[{{Claret} \& {Torres}(2019)}]{claret2019}
{Claret}, A. \& {Torres}, G. 2019, ApJ, 876, 134 (15 pp.)

\bibitem[{Dethero {et~al.}(2024)Dethero, {Pratt}, Vlaykov, {Baraffe}, Guillet,
  {Goffrey}, {Le Saux}, \& Morison}]{Dethero2024}
Dethero, M.-G., {Pratt}, J., Vlaykov, D., {et~al.} 2024, A\&A, 692, A46 (15
  pp.)

\bibitem[{Gough(1977)}]{gough77b}
Gough, D.~O. 1977, ApJ, 214, 196

\bibitem[{Gryanik \& Hartmann(2002)}]{gryanik02b}
Gryanik, V.~M. \& Hartmann, J. 2002, J. Atm. Sci., 59, 2729

\bibitem[{Gryanik {et~al.}(2005)Gryanik, Hartmann, Raasch, \&
  Schr\"oter}]{gryanik05b}
Gryanik, V.~M., Hartmann, J., Raasch, S., \& Schr\"oter, M. 2005, J. Atm. Sci.,
  62, 2632

\bibitem[{Hanjali{\'c} \& Launder(1976)}]{hanjalic76b}
Hanjali{\'c}, K. \& Launder, B.~E. 1976, J. Fluid Mech., 74, 593

\bibitem[{Hansen {et~al.}(2004)Hansen, {Kawaler}, \& Trimble}]{Hansen2004}
Hansen, C.~J., {Kawaler}, S.~D., \& Trimble, V. 2004, Stellar Interiors.
  Physical Principles, Structure, and Evolution., second edition edn.
  (Springer)

\bibitem[{{Hanson} {et~al.}(2024){Hanson}, {Das}, {Mani}, {Hanasoge}, \&
  {Sreenivasan}}]{Hanson2024}
{Hanson}, C.~S., {Das}, S.~B., {Mani}, P., {Hanasoge}, S., \& {Sreenivasan},
  K.~R. 2024, Nat. Astron., 8, 1088

\bibitem[{Houdek \& Dupret(2015)}]{houdek15b}
Houdek, G. \& Dupret, M.-A. 2015, Liv. Rev. Sol. Phys., 12:8, 88 pages

\bibitem[{{K\"aply\"a}(2024)}]{Kaepylae2024}
{K\"aply\"a}, P. 2024, A\&A, 683, A221 (16 pp.)

\bibitem[{{K\"aply\"a}(2025)}]{Kaepylae2025}
{K\"aply\"a}, P. 2025, A\&A, 698, L13 (8 pp.)

\bibitem[{{Kippenhahn} {et~al.}(2012){Kippenhahn}, {Weigert}, \&
  {Weiss}}]{Kippenhahn2012}
{Kippenhahn}, R., {Weigert}, A., \& {Weiss}, A. 2012, {Stellar Structure and
  Evolution}, Astronomy and Astrophysics Library (Springer-Verlag, Berlin
  Heidelberg)

\bibitem[{{Kuhfu{\ss}}(1986)}]{kuhfuss1986}
{Kuhfu{\ss}}, R. 1986, A\&A, 160, 116

\bibitem[{{Kuhfu{\ss}}(1987)}]{kuhfuss1987}
{Kuhfu{\ss}}, R. 1987, PhD thesis, Max-Planck-Institut f\"ur Astrophysik,
  Technische Universit\"at M\"unchen

\bibitem[{Kupka(1999)}]{kupka99b}
Kupka, F. 1999, ApJL, 526, L45

\bibitem[{Kupka(2007)}]{kupka07f}
Kupka, F. 2007, in Symposium, Vol. 239, Convection in Astrophysics, ed.
  F.~Kupka, I.~W. Roxburgh, \& K.~L. Chan, IAU (Cambridge University Press),
  92--94

\bibitem[{{Kupka} {et~al.}(2022){Kupka}, {Ahlborn}, \& {Weiss}}]{Kupka2022}
{Kupka}, F., {Ahlborn}, F., \& {Weiss}, A. 2022, A\&A, 667, A96 (17 pp.)

\bibitem[{Kupka \& Montgomery(2002)}]{kupka02b}
Kupka, F. \& Montgomery, M.~H. 2002, MNRAS, 330, L6

\bibitem[{Kupka \& Muthsam(2007{\natexlab{a}})}]{kupka07d}
Kupka, F. \& Muthsam, H.~J. 2007{\natexlab{a}}, in Symposium, Vol. 239,
  Convection in Astrophysics, ed. F.~Kupka, I.~W. Roxburgh, \& K.~L. Chan, IAU
  (Cambridge University Press), 83--85

\bibitem[{Kupka \& Muthsam(2007{\natexlab{b}})}]{kupka07e}
Kupka, F. \& Muthsam, H.~J. 2007{\natexlab{b}}, in Symposium, Vol. 239,
  Convection in Astrophysics, ed. F.~Kupka, I.~W. Roxburgh, \& K.~L. Chan, IAU
  (Cambridge University Press), 86--88

\bibitem[{{Kupka} \& {Muthsam}(2017)}]{kupka17b}
{Kupka}, F. \& {Muthsam}, H.~J. 2017, Liv. Rev. Comp. Astrophys., 3, 1 (159
  pp.)

\bibitem[{Kupka \& Robinson(2007)}]{kupka07b}
Kupka, F. \& Robinson, F.~J. 2007, MNRAS, 374, 305

\bibitem[{Kupka {et~al.}(2018)Kupka, Zaussinger, \& Montgomery}]{kupka18b}
Kupka, F., Zaussinger, F., \& Montgomery, M.~H. 2018, MNRAS, 474, 4660

\bibitem[{Lesieur(2008)}]{Lesieur08b}
Lesieur, M. 2008, Turbulence in Fluids, 4th edn. (Dordrecht: Springer)

\bibitem[{Li(2012)}]{li12b}
Li, Y. 2012, ApJ, 756, 37

\bibitem[{Losch(2004)}]{losch04b}
Losch, M. 2004, Geophys. Res. Lett., 31, L23301 (4 pp.)

\bibitem[{Millionshchikov(1941)}]{millionshchikov41b}
Millionshchikov, M.~D. 1941, Dokl. Acad. Nauk SSSR, 32, 611

\bibitem[{Mironov(2009)}]{mironov09b}
Mironov, D.~V. 2009, in Lecture Notes in Physics, Vol. 756, Interdisciplinary
  {A}spects of {T}urbulence, ed. W.~Hillebrandt \& F.~Kupka (Berlin: Springer),
  161--221

\bibitem[{Mironov {et~al.}(1999)Mironov, Gryanik, Lykossov, \&
  Zilitinkevich}]{mironov99b}
Mironov, D.~V., Gryanik, V.~M., Lykossov, V.~N., \& Zilitinkevich, S.~S. 1999,
  J. Atm. Sci., 56, 3478

\bibitem[{Montgomery \& Kupka(2004)}]{montgomery04b}
Montgomery, M.~H. \& Kupka, F. 2004, MNRAS, 350, 267

\bibitem[{Mosumgaard {et~al.}(2020)Mosumgaard, J{\o}rgensen, Weiss, Aguirre, \&
  Christensen-Dalsgaard}]{Mosumgaard2020}
Mosumgaard, J., J{\o}rgensen, A., Weiss, A., Aguirre, V., \&
  Christensen-Dalsgaard, J. 2020, MNRAS, 491, 1160

\bibitem[{Muthsam {et~al.}(2010)Muthsam, Kupka, L{\"o}w-Baselli, Obertscheider,
  Langer, \& Lenz}]{Muthsam10a}
Muthsam, H., Kupka, F., L{\"o}w-Baselli, B., {et~al.} 2010, NewA, 15, 460

\bibitem[{Nordlund(1976)}]{Nordlund1976}
Nordlund, {\AA}. 1976, A\&A, 50, 23

\bibitem[{{Nordlund} {et~al.}(2009){Nordlund}, {Stein}, \&
  {Asplund}}]{nordlund09r}
{Nordlund}, {\AA}., {Stein}, R.~F., \& {Asplund}, M. 2009, Liv. Rev. Sol.
  Phys., 6:2, 117 pages

\bibitem[{Pope(2000)}]{pope00b}
Pope, S.~B. 2000, Turbulent {F}lows (Cambridge: Cambridge University Press)

\bibitem[{{Pratt} {et~al.}(2017){Pratt}, {Baraffe}, {Goffrey}, {Constantino},
  {Viallet}, {Popov}, {Walder}, \& {Folini}}]{pratt17a}
{Pratt}, J., {Baraffe}, I., {Goffrey}, T., {et~al.} 2017, A\&A, 604, A125 (15
  pp.)

\bibitem[{Robinson {et~al.}(2003)Robinson, Demarque, Li, Sofia, Kim, Chan, \&
  Guenther}]{robinson03b}
Robinson, F.~J., Demarque, P., Li, L.~H., {et~al.} 2003, MNRAS, 340, 923

\bibitem[{Rosenthal {et~al.}(1999)Rosenthal, Christensen-Dalsgaard, Nordlund,
  Stein, \& Trampedach}]{rosenthal99b}
Rosenthal, C.~S., Christensen-Dalsgaard, J., Nordlund, {\AA}., Stein, R.~F., \&
  Trampedach, R. 1999, ApJ, 351, 689

\bibitem[{Roxburgh(1989)}]{roxburgh89b}
Roxburgh, I.~W. 1989, A\&A, 211, 361

\bibitem[{{Salaris} \& {Cassisi}(2005)}]{Salaris2005}
{Salaris}, M. \& {Cassisi}, S. 2005, Evolution of Stars and Stellar Population
  (Wiley)

\bibitem[{{Schwarzschild}(1906)}]{schwarzschild06r}
{Schwarzschild}, K. 1906, {N}achr. {K}oenigl. {G}esellsch. {W}iss.
  {G}oettingen, {M}ath.-{P}hys. {K}l., 195, 41

\bibitem[{Spada {et~al.}(2021)Spada, Demarque, \& {Kupka}}]{Spada21}
Spada, F., Demarque, P., \& {Kupka}, F. 2021, MNRAS, 504, 3128

\bibitem[{{Spruit}(1997)}]{Spruit1997}
{Spruit}, H.~C. 1997, Mem. della Soc. Astron. Ital., 68, 397

\bibitem[{Stein \& Nordlund(1998)}]{stein98b}
Stein, R.~F. \& Nordlund, {\AA}. 1998, ApJ, 499, 914

\bibitem[{Stellingwerf(1982)}]{stellingwerf82b}
Stellingwerf, R.~F. 1982, ApJ, 262, 330

\bibitem[{Tennekes \& Lumley(1972)}]{tennekes72b}
Tennekes, H. \& Lumley, J.~L. 1972, A {F}irst {C}ourse in {T}urbulence (M.I.T.
  Press)

\bibitem[{Ulrich(1970)}]{Ulrich1970}
Ulrich, R.~K. 1970, Astrophys. Space Sci., 9, 80

\bibitem[{VandenBerg {et~al.}(2006)VandenBerg, Bergbusch, \&
  Dowler}]{VandenBerg2006}
VandenBerg, D.~A., Bergbusch, P.~A., \& Dowler, P.~D. 2006, ApJS, 162, 375

\bibitem[{{Weiss} {et~al.}(2004){Weiss}, {Hillebrandt}, {Thomas}, \&
  {Ritter}}]{cox04b}
{Weiss}, A., {Hillebrandt}, W., {Thomas}, H.-C., \& {Ritter}, H. 2004, Cox \&
  Giuli's Principles of Stellar Structure. Extended Second Edition (Cambridge:
  Cambridge Scientific Publishers)

\bibitem[{Xiong(1978)}]{xiong78b}
Xiong, D.~R. 1978, Chin. Astron., 2, 118

\bibitem[{Xiong(1985)}]{xiong85b}
Xiong, D.~R. 1985, A\&A, 150, 133

\bibitem[{Xiong(1986)}]{xiong86b}
Xiong, D.~R. 1986, A\&A, 167, 239

\bibitem[{Xiong {et~al.}(1997)Xiong, Cheng, \& Deng}]{xiong97b}
Xiong, D.~R., Cheng, Q.~L., \& Deng, L. 1997, ApJS, 108, 529

\bibitem[{Yokoi(2023)}]{Yokoi2023}
Yokoi, N. 2023, atmosphere, 14, 1013 (22 pp.)

\bibitem[{Yokoi {et~al.}(2022)Yokoi, Masada, \& Takiwaki}]{Yokoi2022}
Yokoi, N., Masada, Y., \& Takiwaki, T. 2022, MNRAS, 516, 2718

\bibitem[{Zahn(1991)}]{zahn91b}
Zahn, J.-P. 1991, A\&A, 252, 179

\bibitem[{Zilitinkevich {et~al.}(1999)Zilitinkevich, Gryanik, Lykossov, \&
  Mironov}]{zilitinkevich99b}
Zilitinkevich, S., Gryanik, V.~M., Lykossov, V.~N., \& Mironov, D.~V. 1999, J.
  Atm. Sci., 56, 3463

\end{thebibliography}

\appendix
\section{A variant of the new TOM model for the Kuhfu{\ss} 3-equation model of convection}  \label{app_KF_model} 

To implement the TOM model as defined by Eq.~(\ref{eq_w3_new}), Eq.~(\ref{eq_w2t_homogenized}), and 
Eq.~(\ref{eq_wt2_homogenized}) into the model of \citet{kuhfuss1987}, used in the modification suggested by 
\citet{Kupka2022} and \citet{Ahlborn2022}, Eq.~(\ref{eq_Sw_new}) is rewritten in terms of the total turbulent kinetic 
energy $\omega \equiv K$ by means of  Eq.~(\ref{eq_rewrite_K_w2}). Differences due to isotropy assumptions 
are absorbed into the closure constants such that
\begin{eqnarray}  \label{eq_Somega_new}
 S_{\omega} \! &\!\! = \!\! &\! a_{\omega}\, \left(1 - \frac{2\, \tilde{c}_6\ell \omega^{-1/2} \tilde{N}}{1+ 2\, \tilde{c}_6\ell \omega^{-1/2} \tilde{N}} \right)  
                      + d_{\omega}  \ell\, \omega^{\,-1} \frac{\partial{\omega}}{\partial z} \,\,\, \mathrm{if}\,\, N  > 0,  \nonumber \\
 S_{\omega} \! &\!\! = \!\! &\! a_{\omega} + d_{\omega} \ell\, \omega^{\,-1} \frac{\partial{\omega}}{\partial z} \quad \mathrm{otherwise}.
\end{eqnarray}
Furthermore. $a_{\omega} <  0$, $d_{\omega} < 0$, and $\tilde{c}_6$ have to be adjusted from the values for $a_6$, $d_6$, and $c_6$
provided in the main text. With $\overline{\rho}$ as the mean density the non-local flux $F_{\omega}$ can be computed from
\begin{equation}  \label{eq_omega_new}
   F_{\omega} = \frac{1}{\overline{\rho}}\,\mathrm{div}\left(S_{\omega}\,\overline{\rho}\, \omega^{3/2} \right).
\end{equation}
The \citet{kuhfuss1987} model computes correlations for entropy fluctuations rather than temperature 
fluctuations. It hence requires to calculate the non-local fluxes of $\Pi = \overline{w\,s'}$, 
which is the (radial component of the) flux of entropy fluctuations, closely related to the convective flux, and 
of $\phi = \overline{s'^2}$, the square of local entropy fluctuations. For a well mixed plasma in the
temperature and density region of interest to stellar evolution calculations, the entropy fluctuations correlate
very well with the temperature fluctuations of the fluid. Thus, the non-local fluxes of $\Pi$ and  $\phi$ are suggested
to be obtained from  Eq.~(\ref{eq_w2t_homogenized}) and Eq.~(\ref{eq_wt2_homogenized}) by replacing $\theta$
with $s'$, setting $v'^2=q^2$, and writing their non-local fluxes in terms of $\omega$ instead of $w$ as follows:
\begin{equation}   
F_{\Pi} = \frac{1}{\overline{\rho}}\,\mathrm{div}\left( \overline{\rho}\,\overline{v'^2 s'}\right) 
            =  \frac{1}{\overline{\rho}}\,\mathrm{div}\left( \overline{\rho}\, \left(a_{\Pi}\, \Pi\, \omega^{1/2}\, S_{\omega} 
                                  +d_{\Pi}\, \ell\, \omega^{1/2}\, \frac{\partial{\Pi}}{\partial  z}\right) \right)  \label{eq_w2t_with_omega}
\end{equation}
and
\begin{equation}
F_{\phi} = \frac{1}{\overline{\rho}}\,\mathrm{div}\left( \overline{\rho}\,\overline{v' s'^2}\right) 
              =  \frac{1}{\overline{\rho}}\,\mathrm{div}\left( \overline{\rho}\, \left(a_{\phi}\,\Pi^2\, \omega^{-1/2}\, S_{\omega} 
                                        +d_{\phi}\, \ell\, \omega^{1/2}\,  \frac{\partial{\phi}}{\partial  z}\right) \right). \label{eq_wt2_with_omega} 
\end{equation}
Again the constants $a_{\Pi} >  0$ and $d_{\Pi} < 0$ as well as $a_{\phi} >  0$ and $d_{\phi} < 0$
have to be adjusted from the values for $a_1$ and $d_4$ as well as $a_5$ and $d_5$ given in the main text.

\section{3D RHD simulations of solar convection with closed boundary conditions}  \label{app_comp_FJR} 

\begin{figure}[ht]
\centering
\includegraphics[width=0.45\textwidth]{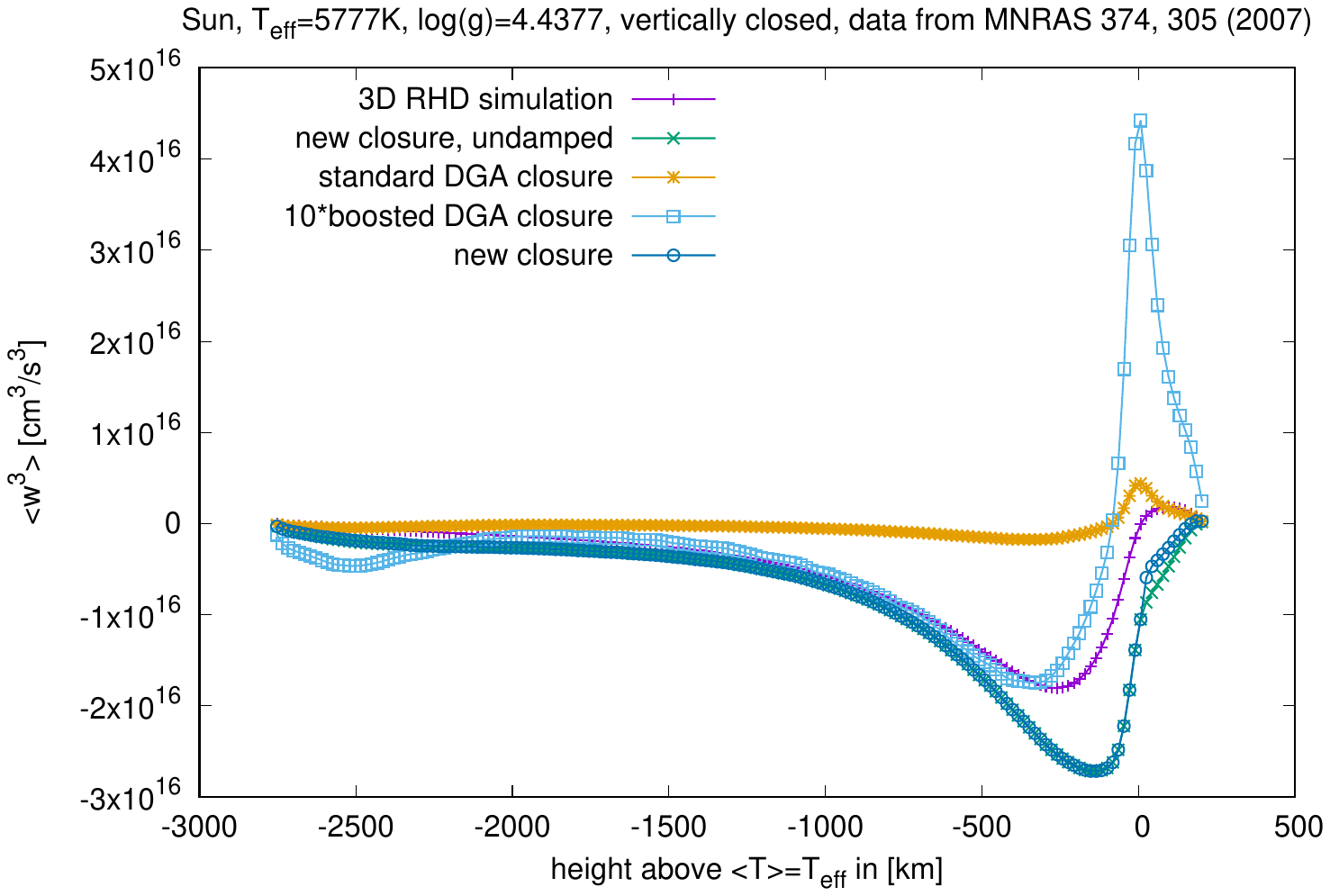}
\includegraphics[width=0.45\textwidth]{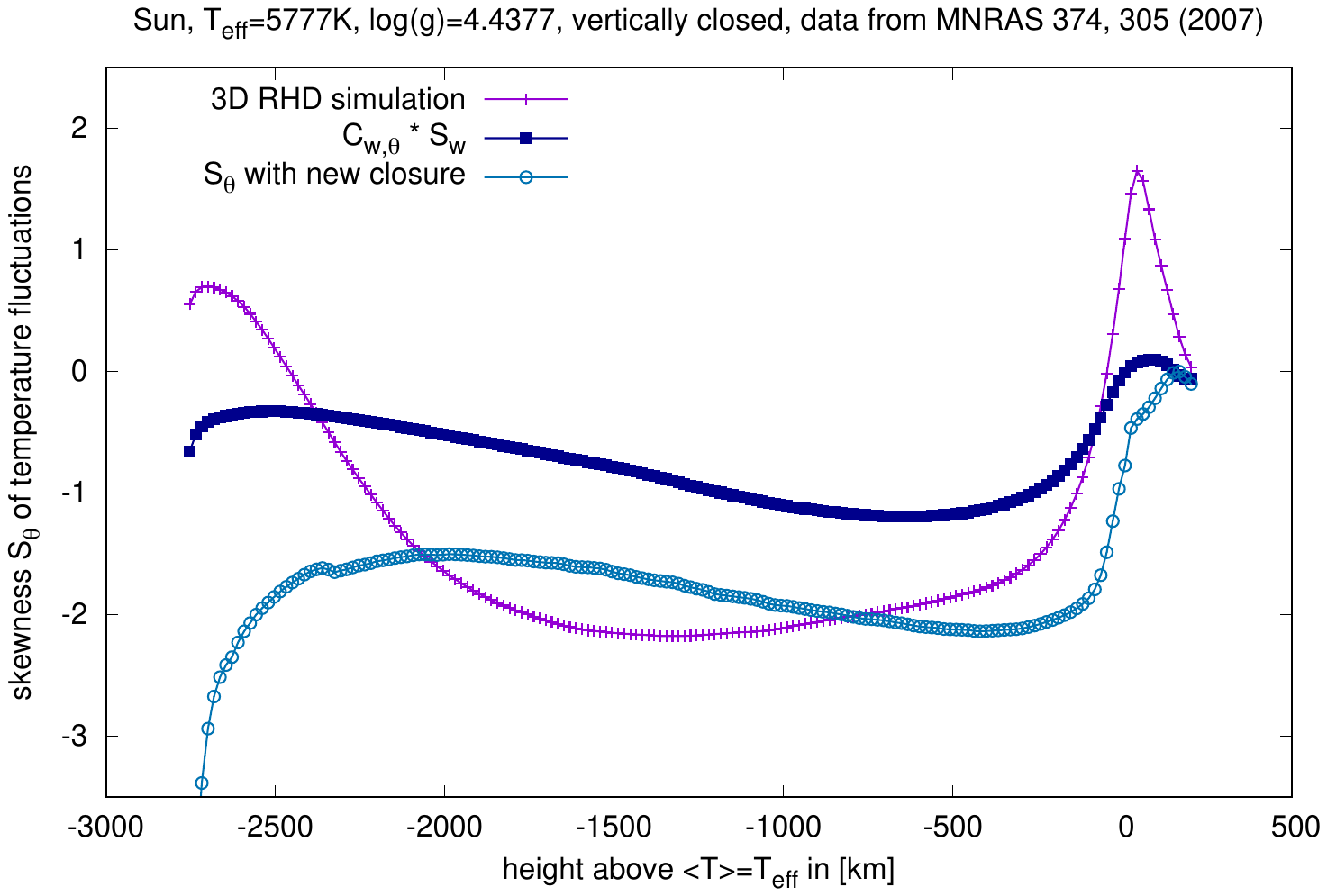}
\caption{Further investigation of the models for the TOM $\overline{w^3}$ and the skewness of temperature $S_{\theta}$.
              The upper panel shows $\overline{w^3}$ and the lower one $S_{\theta}$. Closure models as well as colour and 
              line style coding are the same as used for Fig.~\ref{Fig2}, though for the case of the 3D RHD simulation of solar surface 
              convection with closed vertical boundary conditions. 
\label{FigB1}
}
\end{figure}

\begin{figure}[t]
\centering
\includegraphics[width=0.45\textwidth]{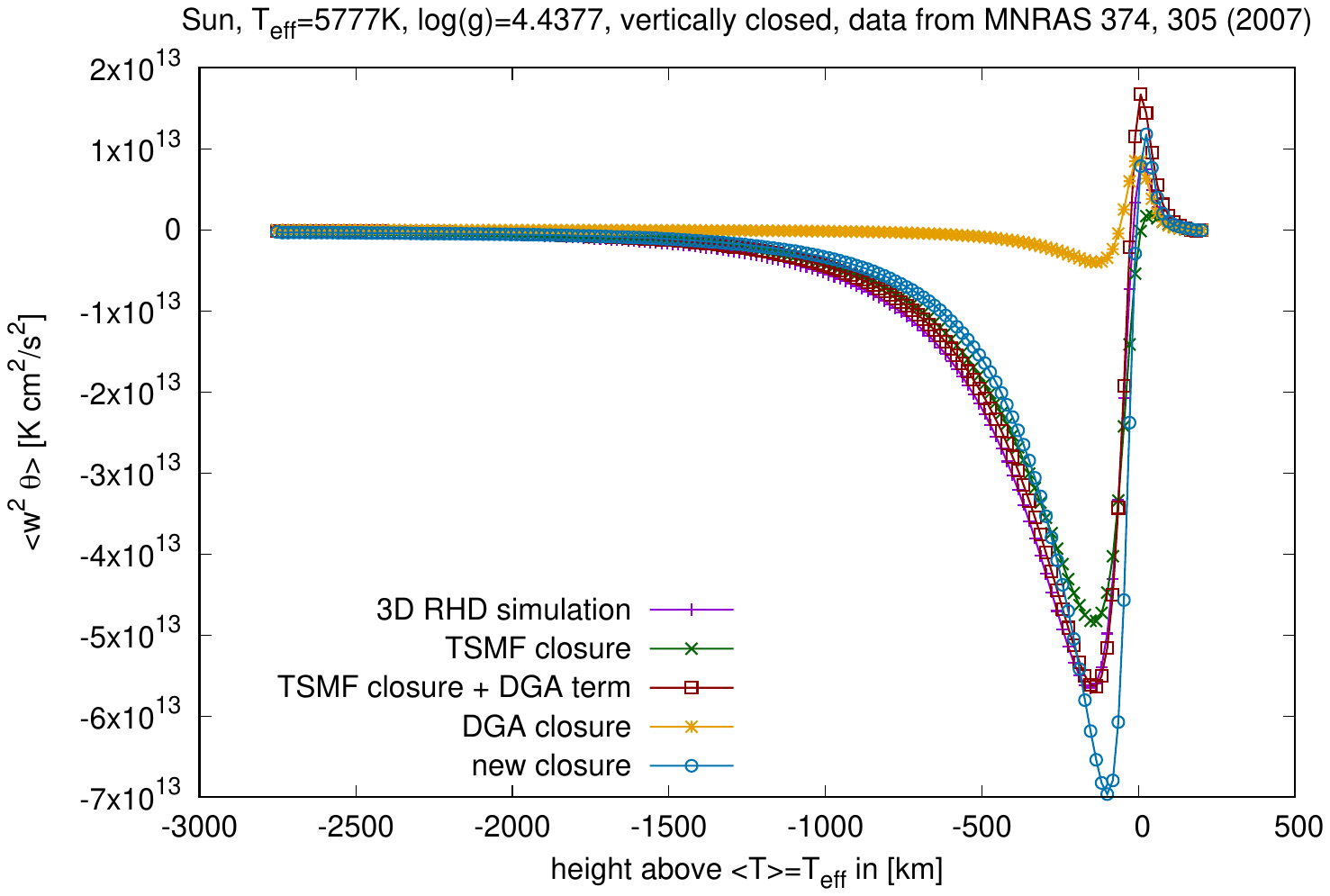}
\includegraphics[width=0.45\textwidth]{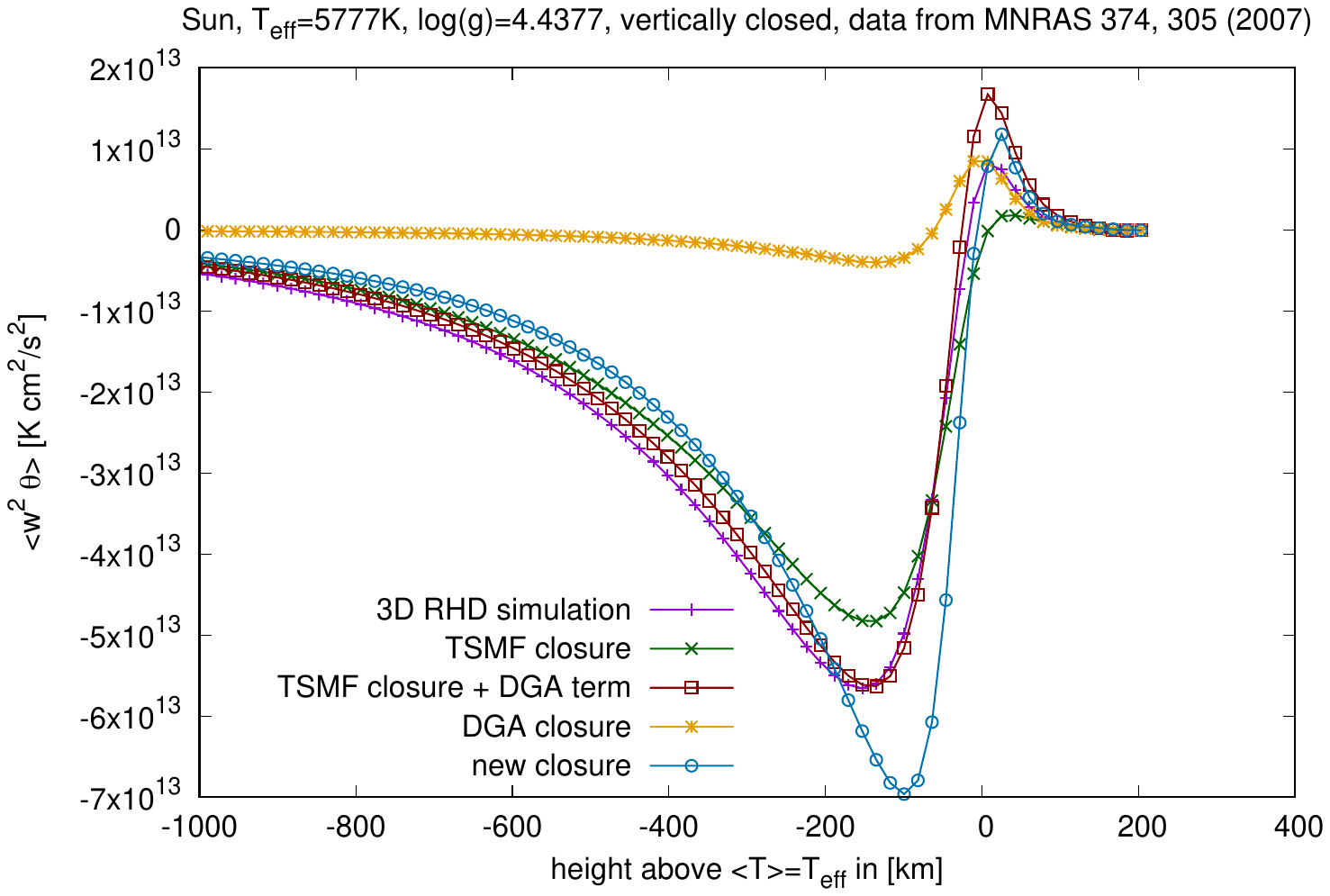}
\caption{Probing the TOM $\overline{w^2\theta}$ with 3D RHD solar simulation data 
              for closed vertical boundary conditions. Details on the line style and colour coding
              are explained in the captions to Fig.~\ref{Fig2} and Fig.~\ref{Fig3}. The upper panel shows the full vertical
              depth range, the lower one  a zoom into the upper 40\% of that range.
\label{FigB2}
}
\end{figure}

\begin{figure}[t]
\centering
\includegraphics[width=0.45\textwidth]{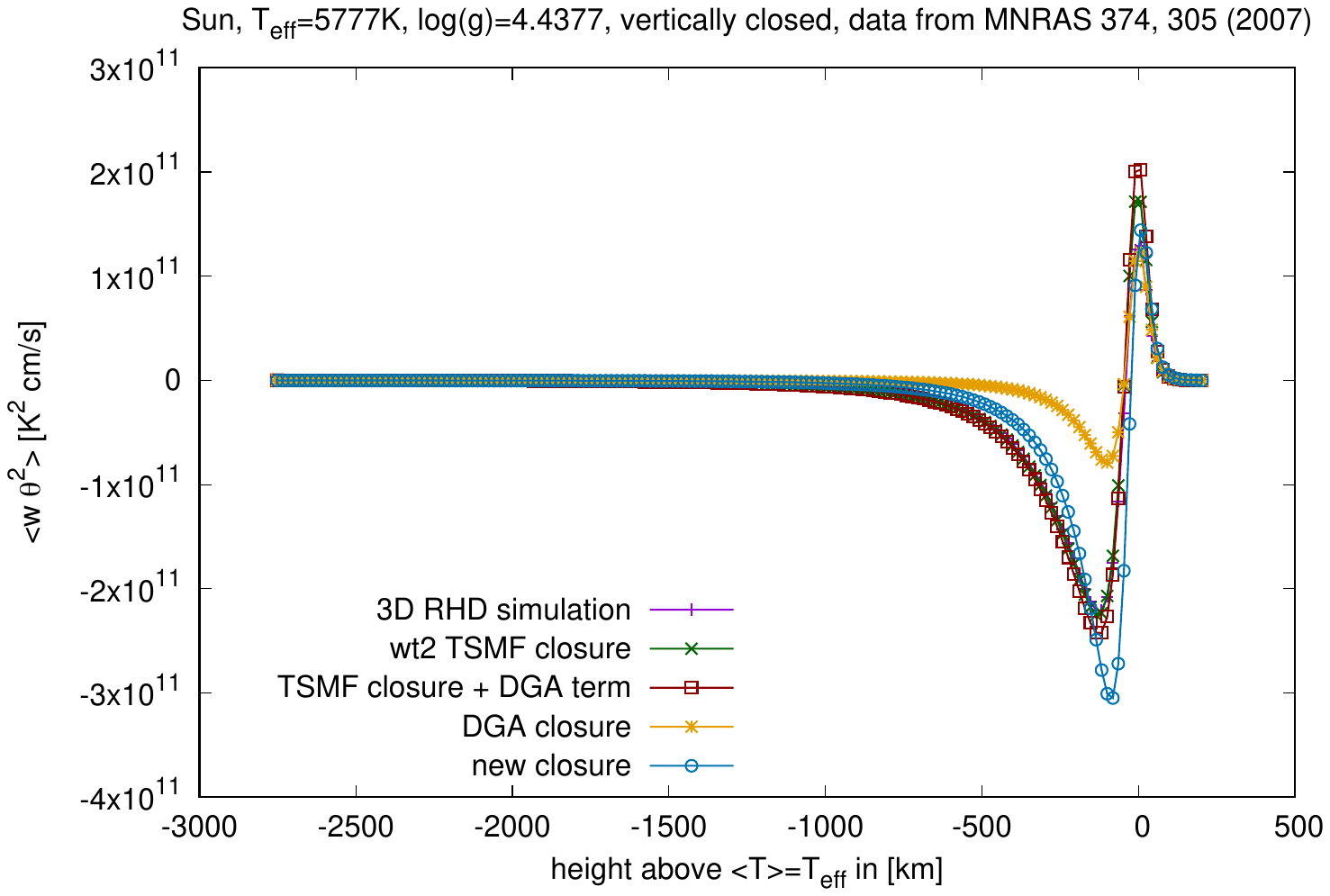}
\includegraphics[width=0.45\textwidth]{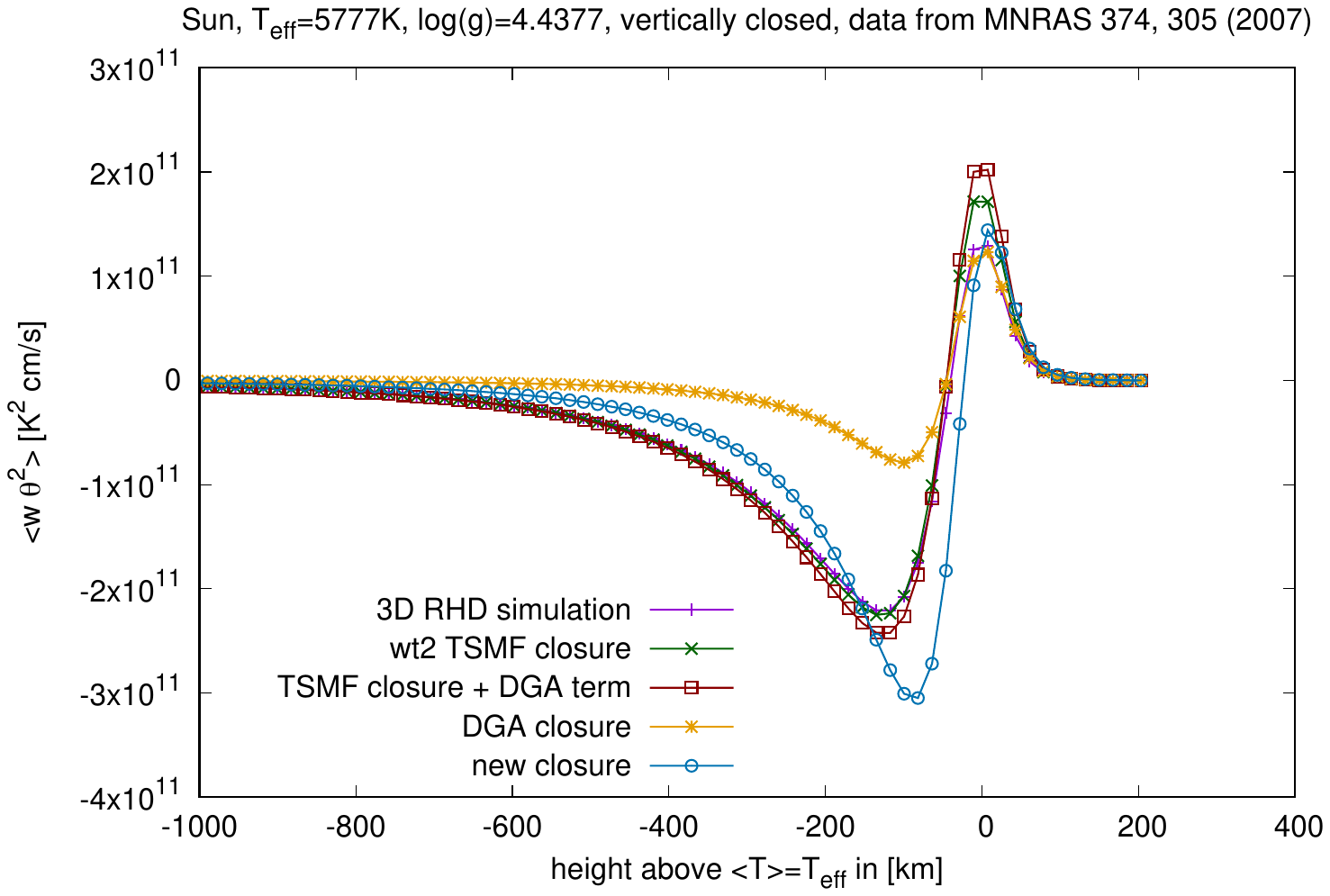}
\caption{Probing the TOM $\overline{w\theta^2}$ with 3D RHD solar simulation data 
              for closed vertical boundary conditions. Details on the line style and colour coding
              can be found in the captions to Fig.~\ref{Fig2} and Fig.~\ref{Fig3}. The upper panel shows the full vertical
              depth range, the lower one  a zoom into the upper 40\% of that range.
\label{FigB3}
}
\end{figure}

Third order and other higher order moments of the fluctuations $w$ and $\theta$
of the basic variables velocity and temperature around an ensemble averaged 
mean state are sensitive to boundary conditions assumed for the spatial domain
chosen for 3D RHD numerical simulations. To quantify their impact the tests of
the new TOM model were also performed for a simulation of solar granulation that
assumed a closed vertical boundary in the photosphere and a closed boundary
with a small artificial heating zone inside the quasi-adiabatic envelope 
\citep{robinson03b,kupka07b}. A direct comparison with a 3D RHD simulation
of solar granulation which allows in- and outflow through the vertical boundary
\citep{belkacem19b} was given in Fig.~\ref{Fig1}. Due to the solid plates of
the 3D RHD simulation presented in \citet{kupka07b} $S_w$ did not 
feature a plateau in the interior of the quasi-adiabatic convective envelope
of the Sun. At a distance of almost three local pressure scale heights
from the bottom of the simulation domain $|S_w|$ already begins to drop
until it finally diverges towards the bottom. The DGA initially appears
to follow this behaviour at least if the vertical turbulent momentum diffusivity
were boosted by a factor of ten, but it diverges also much farther away
from the lower boundary. A compromise could have been achieved
by lowering $|a_6|$ and increasing $|d_6|$ in the new TOM model, though 
neither of the models was developed to account for the ``solid plate scenario''. While
the latter is of little interest to solar physics, an important lesson can be learned
if the new TOM model were applied to the PBL of the Earth: the solid terrestrial
surface could influence the vertical velocity field such that a constant 
skewness component could not be disentangled from a turbulent component.
Indeed, in the PBL $S_w$ slightly increases away from the bottom boundary
(see \citealt{gryanik02b} and \citealt{cheng05b}) with no observable plateau,
just as in the DA white dwarf simulation \citep{kupka18b} discussed in Fig.~\ref{Fig4}
in Sect.~\ref{sect_tests}. The upper boundary of the simulation of \citet{kupka07b}
is also responsible for $\overline{w^3}$ changing sign near the top.

The upper panel of Fig.~\ref{FigB1} demonstrates that these peculiarities 
caused by the closed boundary conditions can barely be noticed 
if $\overline{w^3}$ is considered. A very good match could have been achieved 
by slightly lowering $|a_6|$ and doubling $c_6$ which would have lowered 
the contribution of the non-local  (algebraic) component. The influence of
the vertical boundary conditions is similar, if $S_{\theta}$ is considered 
instead of $S_w$ (Fig.~\ref{FigB1}, lower panel).

Figs.~\ref{FigB2} and~Fig.~\ref{FigB3} show that similar to $\overline{w^3}$
illustrated in Fig.~\ref{FigB1}, the TOMs $\overline{w^2\theta}$ and 
$\overline{w\theta^2}$ computed from the 3D RHD simulations were 
reproduced quite well by the new TOM model without any parameter
readjustments. The discrepancies between direct computations of $S_w$
and $S_{\theta}$ from the simulation and their evaluation from a model with 
simulation data for second order moments and mean quantities as input
are similar to the case with open vertical boundary conditions
illustrated in Fig.~\ref{Fig3} of Sect.~\ref{sect_tests}. This underlines
the central role of a detailed modelling of the skewnesses $S_w$
and $S_{\theta}$ for future improvements of the new TOM model.

It is interesting to note that also for this case each of the gradients
of $\overline{w^2}$, $\overline{w\theta}$, and $\overline{\theta^2}$
has a similar shape, as it has for the cases discussed in Sect.~\ref{sect_tests}.
This shape prevents a proper match of $\overline{w^3}$ and $S_w$ 
as well as $\overline{w^2\theta}$ and $\overline{w\theta^2}$ in the case 
of open upper boundary conditions and the same was found to hold
for the  3D RHD simulation with closed vertical boundary conditions.

\section{The computation of the flux of (turbulent) kinetic energy}  \label{app_fkin} 

\begin{figure}[t]
\centering
\includegraphics[width=0.90\columnwidth]{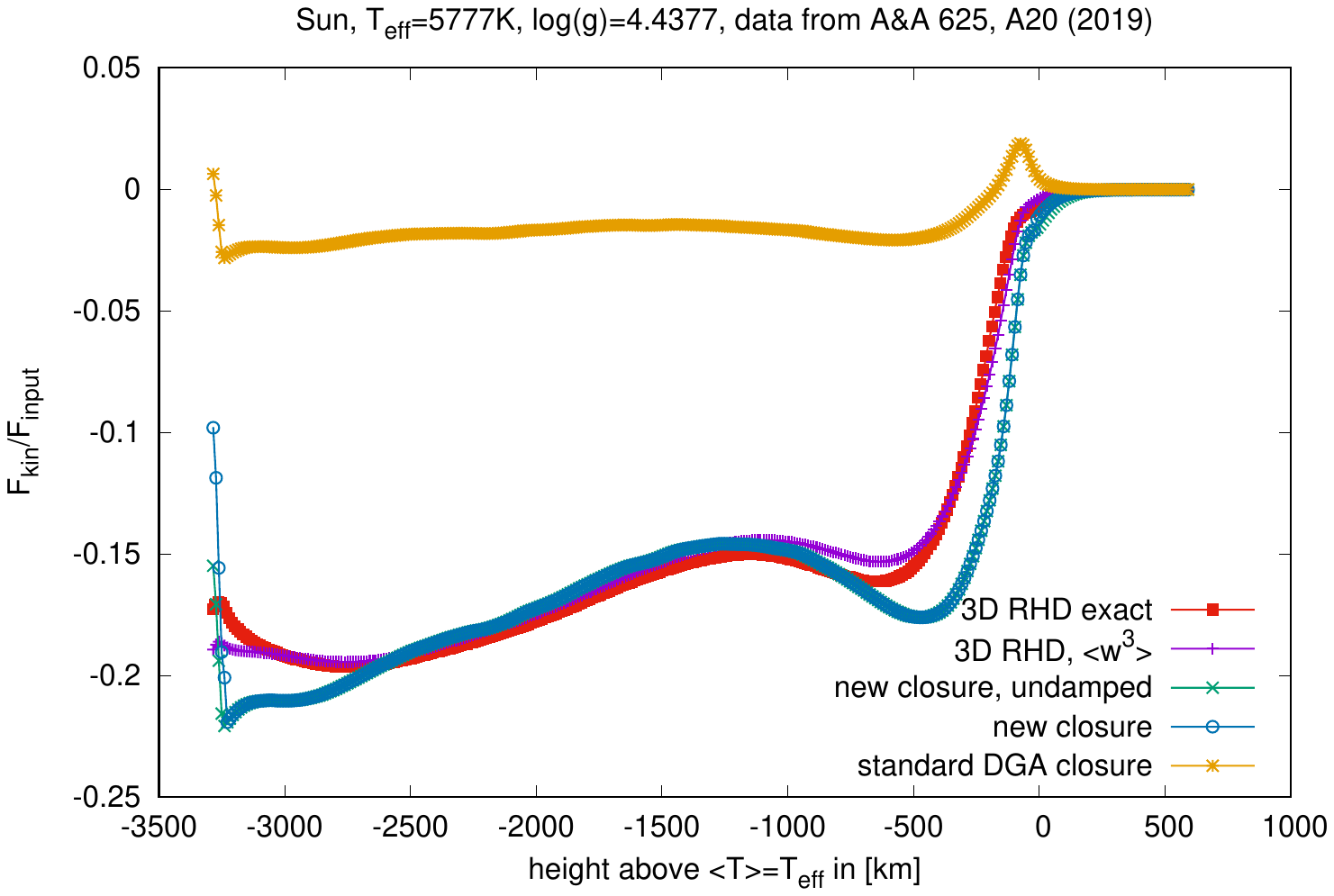}
\includegraphics[width=0.90\columnwidth]{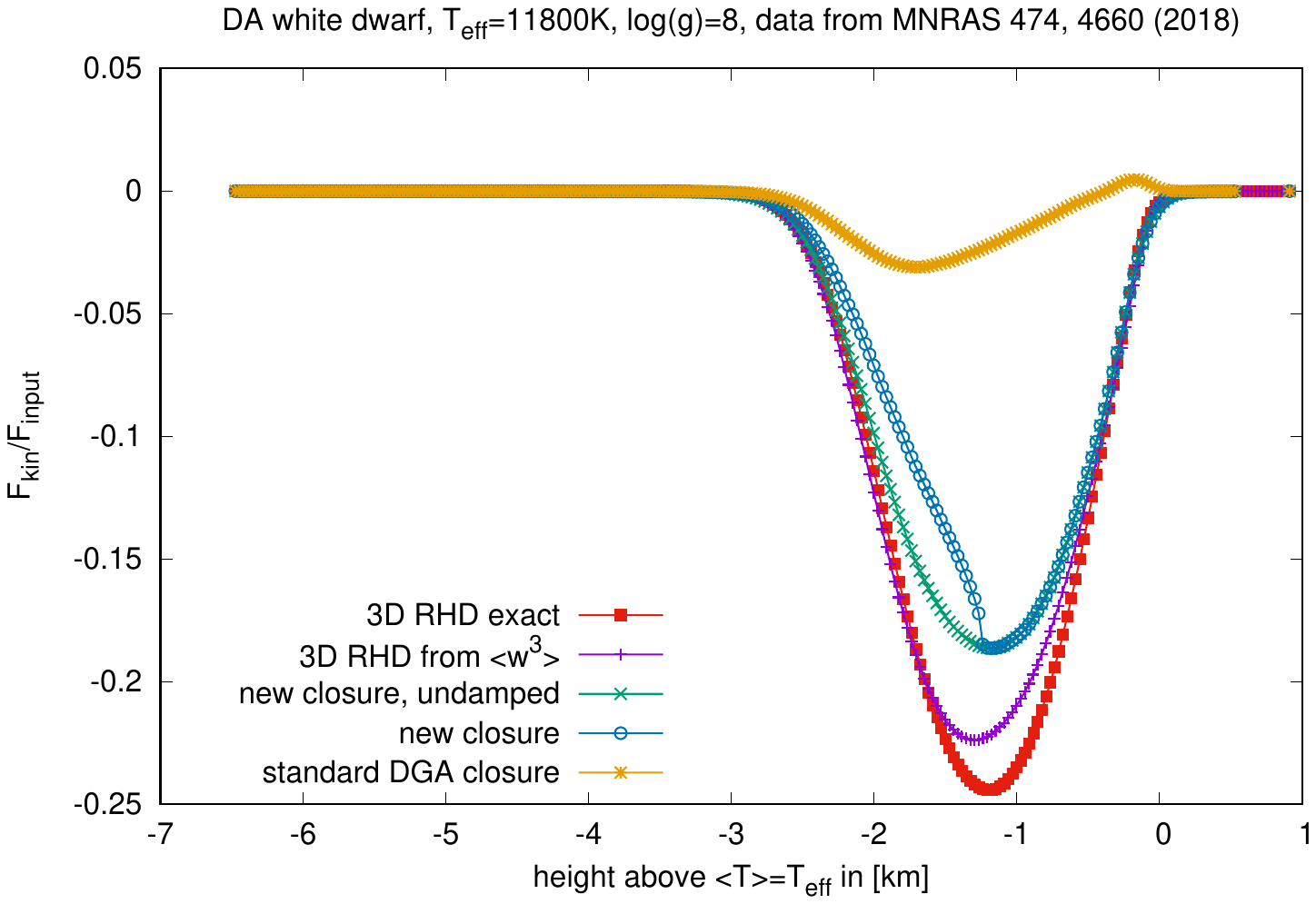}
\includegraphics[width=0.90\columnwidth]{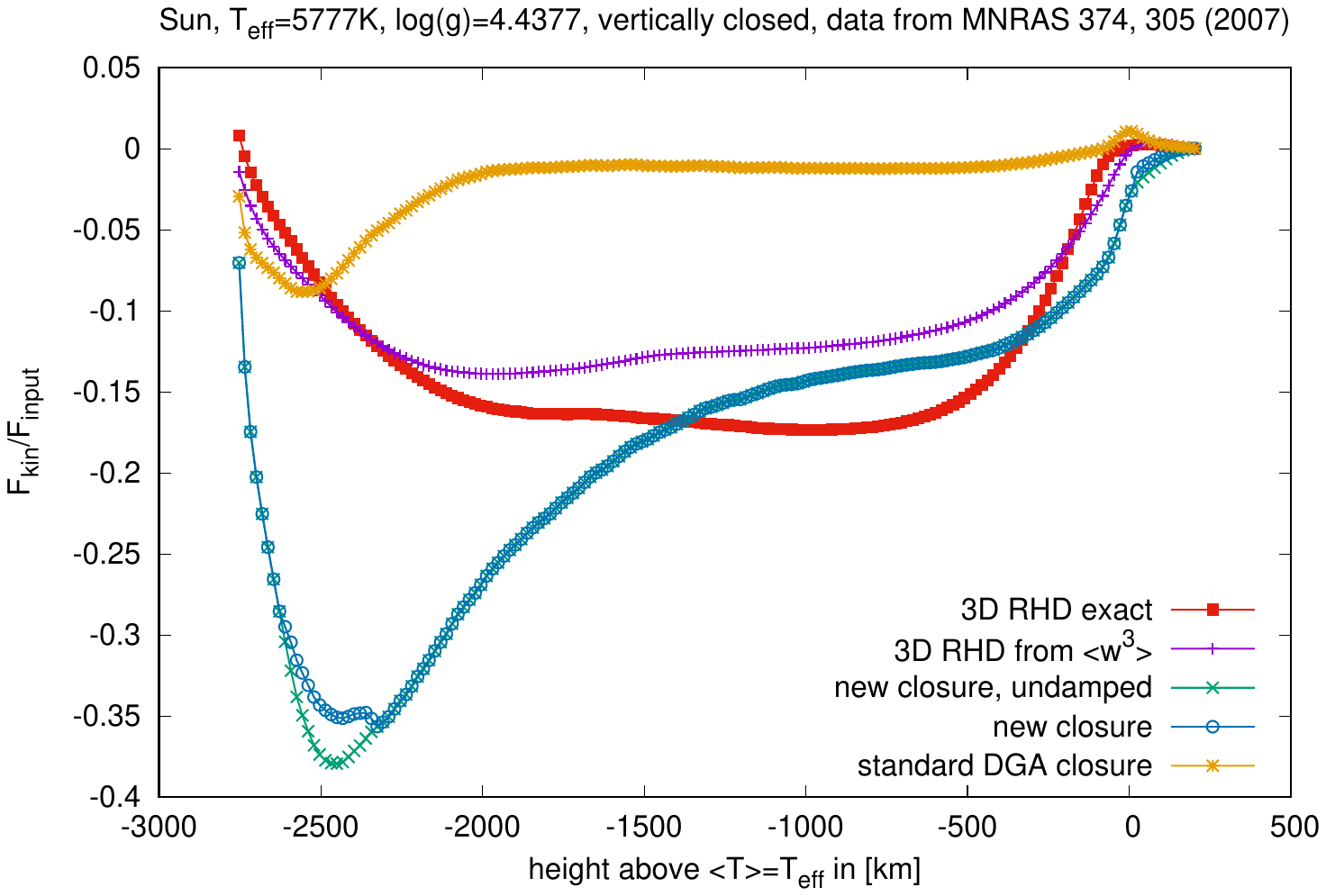}
\caption{Predictions of the flux of (turbulent) kinetic energy in three different 3D RHD simulations
              (upper panel: solar case, middle panel: DA white dwarf, lower panel: solar case with
              solid plate at the bottom). The same colour and
              linestyle coding is used as for Fig.~\ref{Fig1}. For the models and for the computation
              from only the vertical component of the kinetic energy flux an identical scaling factor of $2/3$ was
              applied (see text in Appendix~\ref{app_fkin}). The evaluation of the exact expression for this 
              quantity from the 3D RHD simulations is shown in bright red with data points denoted by filled squares.              
\label{FigC1}
}
\end{figure}

An important physical quantity for the modelling of stellar evolution is the flux
of turbulent kinetic energy, $F_{\rm kin}$, which should enter into the flux balance 
of the luminosity equation of stellar evolution models \citep{canuto93b}. As a purely 
non-local quantity it cannot be computed within the framework of the mixing length theory
of  convection. In spite of its crucial role to allow for overshooting in first place 
\citep{canuto93b} $F_{\rm kin}$ is not necessarily accounted for
in the luminosity equation of stellar evolution codes. Fig.~\ref{FigC1} can also be
used to justify that decision: if the TOMs in the non-local convection model are
computed from the DGA, $|F_{\rm kin}|$ barely exceeds 2\% of the total energy
flux and it might be argued for stability reasons that such a small quantity does
not have to be accounted for. In such cases the non-local fluxes in the second order
moment equations still affect the computation of the convective enthalpy flux $|F_{\rm conv}|$
and thus the temperature gradient (the small amount given by $|F_{\rm kin}|$ is
``absorbed'' into $F_{\rm conv}$ in such models). As can be seen from Fig.~\ref{FigC1}, too,
the assumption of a small $|F_{\rm kin}|$ no longer holds once this quantity is directly
computed from 3D RHD numerical simulations of convection at the stellar surface and 
the upper part of the stellar envelope. The exact evaluation of $F_{\rm kin}$ is shown
there as a bright red line with filled squares. It stays in the range of $-15\%$ to $-20\%$
of the total (actually input) flux throughout the quasi-adiabatic part of the convection
zone in the solar case (upper panel). A solid boundary inside such a zone necessarily forces
$F_{\rm kin}$ back to zero (bottom panel), although even in this case remains in
this range except for the lowermost pressure scale height. For the 3D RHD simulation
of the DA white dwarf even a value of $-25\%$ is attained and $F_{\rm kin}$ remains
in the range of $-20\%$ to $-5\%$ throughout the countergradient layer. This is no 
longer a small contribution and an accurate prediction of the temperature structure 
in these classes of stars requires to account for its contribution to the energy flux budget.

When challenging TOM models to predict $F_{\rm kin}$, the first problem to be solved 
is the computation of the triple correlation for the flux of the total kinetic energy, $\overline{q^2 w}$.
For Fig.~\ref{FigC1} this was done by testing whether $F_{\rm kin} =  (1/2) \, \overline{\rho\, q^2  w}$
could be estimated from $F_{\rm kin} \approx c_{\rm Fkin}\, \overline{\rho}\, \overline{w^3}$. 
Fig.~\ref{FigC1} demonstrates that the latter is sufficiently accurate for all cases considered
by simply setting $c_{\rm Fkin} =  2/3$ (see also \citealt{chan96b}). This is likely due to a rather 
constant anisotropy ratio between the total and the vertical (turbulent) kinetic energy. Deviations 
from that constancy can be neglected at the stellar surface due to the rapid drop of density
which makes the contribution of $F_{\rm kin}$ to the total flux negligibly small. The
same procedure was hence applied to estimate $F_{\rm kin}$ from the different models
for $\overline{w^3}$. Again, the new TOM model performs very well except near the
bottom of the ``artificially'' vertically closed solar surface convection simulation shown
in the bottom panel of Fig.~\ref{FigC1}. If the new TOM model is used in non-local
stellar convection models, $F_{\rm kin}$ should hence be accounted for in the 
energy flux budget. It can safely be neglected, if the DGA model is used, since the
latter completely underestimates this quantity, as already mentioned further above.

\section{Arguments based on plume models}  \label{app_plume} 

Plume models as developed in \citet{canuto98b}, \citet{canuto07b}, \citet{canuto09b},
and also \citet{belkacem06b} provide additional arguments for chosing the 
contribution to $S_w$ created by cooling in a narrow boundary layer as a constant,
Eq.~(\ref{eq_Sw_CS96_added}), and scale $\overline{w^3}$
proportional to $(\overline{w^2})^{3/2}$ as in Eq.~(\ref{eq_w3_CS96_added}). 
As discussed in \citet{canuto98b} Eq.~(\ref{eq_sigma}) can easily be rewritten into 
$S_w = (1-2\,\sigma)/\sqrt{\sigma-\sigma^2}$ to yield $S_w$ as a function of (upflow) area 
$\sigma$. For an upflow area fraction of 0.79 (and a downflow area fraction of 0.21) a value of
$S_w=-\sqrt{2} \approx a_6$ is obtained (the choice of sign for $S_w$ when solving this quadratic
equation is straightforward). From an inspection of 3D RHD of solar granulation it follows
\citep{belkacem06b} that values of $S_w$ in the range of $-1.75$ to $-1.5$ are correlated with
values of $\sigma$  in the range of 0.67 to 0.63, whence  Eq.~(\ref{eq_sigma}) overestimates
$\sigma$ (or underestimates $|S_w|$). This can be attributed, among others, to the
averaging procedure used in the two-delta function plume model (see \citealt{belkacem06b} 
for details). To account for such effects would require a dynamical equation for $S_w$ 
and result in a significantly more complex model.  Eq.~(\ref{eq_prop_t3}) for $S_{\theta}$ 
cannot be as easily motivated from the family of two-delta function plume models
since they do not distinguish between $S_w$ and $S_{\theta}$ as required from
3D RHD solar granulation simulations \citep{belkacem06b}. The TSMF model of
\citet{gryanik02b} does so by construction which also allows support for the
derivation of Eq.~(\ref{St_closure}). The detailed dependencies between up- and 
downflows with hot and cold flows were simplified in Sect.~\ref{sect_Stheta}, however,
because this would have considerably increased the complexity of the model.

\end{document}